\documentclass[journal]{IEEEtran}
\usepackage{amsmath,amssymb,amsthm,amsfonts}
\newtheorem{theorem}{Theorem}[section]
\newtheorem{corollary}{Corollary}[section]

\usepackage{graphicx}
\usepackage{xcolor}
\usepackage{pgfplots}
\usepgfplotslibrary{patchplots}  
\usepackage{pgfplotstable}  
\pgfplotsset{compat=1.17}
\usepackage{tikz}                

\usepackage{caption}
\usepackage{subcaption}
\usepackage{cite}
\usepackage[nolist]{acronym}

\makeatletter
\def\subsection{\@startsection{subsection}{2}{\z@}
  {1.0ex plus 0.2ex minus 0.2ex}
  {0.6ex plus 0.1ex}
  {\normalfont\normalsize\itshape}}
\makeatother
	\acrodef{5G-NR}{5G New Radio}
	\acrodef{3GPP}{3rd Generation Partnership Project}
	\acrodef{AC}{address coding}
	\acrodef{ACF}{autocorrelation function}
	\acrodef{ACR}{autocorrelation receiver}
	\acrodef{ADC}{analog-to-digital converter}
	\acrodef{AIC}[AIC]{Akaike information criterion}
	\acrodef{aric}[ARIC]{asymmetric restricted isometry constant}
	\acrodef{arip}[ARIP]{asymmetric restricted isometry property}
	
	\acrodef{ARQ}{automatic repeat request}
	\acrodef{AUB}{asymptotic union bound}
	\acrodef{AWGN}{additive white Gaussian noise}
	\acrodef{CRLB}{Cramér-Rao lower bound} 
	\acrodef{RSS}{received signal strength} 
	
	\acrodef{APSK}[PSK]{asymmetric PSK} 
	
	\acrodef{waric}[AWRICs]{asymmetric weak restricted isometry constants}
	\acrodef{warip}[AWRIP]{asymmetric weak restricted isometry property}
	\acrodef{BCH}{Bose, Chaudhuri, and Hocquenghem}        
	\acrodef{BCHC}[BCHSC]{BCH based source coding}
	\acrodef{BEP}{bit error probability}
	\acrodef{BFC}{block fading channel}
	\acrodef{BG}[BG]{Bernoulli-Gaussian}
	\acrodef{BGG}{Bernoulli-Generalized Gaussian}
	\acrodef{BPAM}{binary pulse amplitude modulation}
	\acrodef{BPDN}{Basis Pursuit Denoising}
	\acrodef{BPPM}{binary pulse position modulation}
	\acrodef{BPSK}{binary phase shift keying}
	\acrodef{BPZF}{bandpass zonal filter}
	\acrodef{BSC}{binary symmetric channels}              
	\acrodef{BU}[BU]{Bernoulli-uniform}
	\acrodef{BER}{bit error rate}
	\acrodef{BS}{base station}

	\acrodef{CP}{Cyclic Prefix}
	\acrodef{CDF}{cumulative distribution function}
	\acrodef{CCDF}{complementary cumulative distribution function}
	\acrodef{CD}{cooperative diversity}
	
	\acrodef{CDMA}{Code Division Multiple Access}
	\acrodef{ch.f.}{characteristic function}
	\acrodef{CIR}{channel impulse response}
	\acrodef{cosamp}[CoSaMP]{compressive sampling matching pursuit}
	\acrodef{CR}{cognitive radio}
	\acrodef{cs}[CS]{compressed sensing}                   
	\acrodef{CSI}{channel state information}
	\acrodef{CCSDS}{consultative committee for space data systems}
	\acrodef{CC}{convolutional coding}
	\acrodef{Covid19}[COVID-19]{Coronavirus disease}

	\acrodef{DAA}{detect and avoid}
	\acrodef{DAB}{digital audio broadcasting}
	\acrodef{DCT}{discrete cosine transform}
	\acrodef{dft}[DFT]{discrete Fourier transform}
	\acrodef{DR}{distortion-rate}
	\acrodef{DS}{direct sequence}
	\acrodef{DS-SS}{direct-sequence spread-spectrum}
	\acrodef{DTR}{differential transmitted-reference}
	\acrodef{DVB-H}{digital video broadcasting\,--\,handheld}
	\acrodef{DVB-T}{digital video broadcasting\,--\,terrestrial}
	\acrodef{DL}{downlink}
	\acrodef{DSSS}{Direct Sequence Spread Spectrum}
	\acrodef{DFT-s-OFDM}{Discrete Fourier Transform-spread-Orthogonal Frequency Division Multiplexing}
	\acrodef{DAS}{distributed antenna system}
	\acrodef{DNA}{Deoxyribonucleic Acid}
	
	\acrodef{EC}{European Commission}
	\acrodef{EED}[EED]{exact eigenvalues distribution}
	\acrodef{EIRP}{Equivalent Isotropically Radiated Power}
	\acrodef{ELP}{equivalent low-pass}
	\acrodef{eMBB}{enhanced mobile broadband}
	\acrodef{EMF}{electric and magnetic fields}
	\acrodef{EU}{European union}

	\acrodef{FC}[FC]{fusion center}
	\acrodef{FCC}{Federal Communications Commission}
	\acrodef{FEC}{forward error correction}
	\acrodef{FFT}{fast Fourier transform}
	\acrodef{FH}{frequency-hopping}
	\acrodef{FH-SS}{frequency-hopping spread-spectrum}
	\acrodef{FSsmall}[FS]{frame synchronization}  
	\acrodef{FDMA}{Frequency Division Multiple Access}    

	\acrodef{GA}{Gaussian approximation}
	\acrodef{GF}{Galois field }
	\acrodef{GG}{Generalized-Gaussian}
	\acrodef{GIC}[GIC]{generalized information criterion}
	\acrodef{GLRT}{generalized likelihood ratio test}
	\acrodef{GPS}{Global Positioning System}
	\acrodef{GMSK}{Gaussian minimum shift keying}
	\acrodef{GSMA}{Global System for Mobile communications Association}
	
	\acrodef{HAP}{high altitude platform}

	\acrodef{IDR}{information distortion-rate}
	\acrodef{IFFT}{inverse fast Fourier transform}
	\acrodef{iht}[IHT]{iterative hard thresholding}
	\acrodef{i.i.d.}{independent, identically distributed}
	\acrodef{IoT}{internet of things}                      
	\acrodef{IR}{impulse radio}
	\acrodef{lric}[LRIC]{lower restricted isometry constant}
	\acrodef{lrict}[LRICt]{lower restricted isometry constant threshold}
	\acrodef{ISI}{intersymbol interference}
	\acrodef{ITU}{International Telecommunication Union}
	\acrodef{ICNIRP}{International Commission on Non-Ionizing Radiation Protection}
	\acrodef{IEEE}{Institute of Electrical and Electronics Engineers}
	\acrodef{ICES}{IEEE international committee on electromagnetic safety}
	\acrodef{IEC}{International Electrotechnical Commission}
	\acrodef{IARC}{International Agency on Research on Cancer}
	\acrodef{IS-95}{Interim Standard 95}
	
	\acrodef{LEO}{low earth orbit}
	\acrodef{LF}{likelihood function}
	\acrodef{LLF}{log-likelihood function}
	\acrodef{LLR}{log-likelihood ratio}
	\acrodef{LLRT}{log-likelihood ratio test}
	\acrodef{LoS}{line-of-sight}
	\acrodef{LRT}{likelihood ratio test}
	\acrodef{wlric}[LWRIC]{lower weak restricted isometry constant}
	\acrodef{wlrict}[LWRICt]{LWRIC threshold}
	\acrodef{LPWAN}{low power wide area network}
	\acrodef{LoRaWAN}{Low power long Range Wide Area Network}
	\acrodef{NLOS}{non-line-of-sight}
	
	\acrodef{MB}{multiband}
	\acrodef{MC}{multicarrier}
	\acrodef{MDS}{mixed distributed source}
	\acrodef{MF}{matched filter}
	\acrodef{m.g.f.}{moment generating function}
	\acrodef{MI}{mutual information}
	\acrodef{MIMO}{multiple-input multiple-output}
	\acrodef{MISO}{multiple-input single-output}
	\acrodef{ML}[ML]{maximum likelihood}
	\acrodef{MMSE}{minimum mean-square error}
	\acrodef{MMV}{multiple measurement vector}
	\acrodef{M-PSK}[${M}$-PSK]{$M$-ary phase shift keying}                       
	\acrodef{M-APSK}[${M}$-PSK]{$M$-ary asymmetric PSK} 
	
	\acrodef{M-QAM}[$M$-QAM]{$M$-ary quadrature amplitude modulation}
	\acrodef{MRC}{maximal ratio combiner}                  
	\acrodef{maxs}[MSO]{maximum sparsity order}                                      
	\acrodef{M2M}{machine to machine}                                                
	\acrodef{MUI}{multi-user interference}
	\acrodef{mMTC}{massive machine type communications}      
	\acrodef{mm-wave}{millimeter-wave}
	\acrodef{MP}{mobile phone}
	\acrodef{MPE}{maximum permissible exposure}
	\acrodef{MAC}{media access control}
	\acrodef{NB}{narrowband}
	\acrodef{NBI}{narrowband interference}
	\acrodef{NLA}{nonlinear sparse approximation}
	\acrodef{NLoS}{non-line of sight}
	\acrodef{NTIA}{National Telecommunications and Information Administration}
	\acrodef{NTP}{National Toxicology Program}
	\acrodef{NHS}{National Health Service}

	\acrodef{OC}{optimum combining}                             
	\acrodef{OC}{optimum combining}
	\acrodef{ODE}{operational distortion-energy}
	\acrodef{ODR}{operational distortion-rate}
	\acrodef{OFDM}{orthogonal frequency-division multiplexing}
	\acrodef{omp}[OMP]{orthogonal matching pursuit}
	\acrodef{OSMP}[OSMP]{orthogonal subspace matching pursuit}
	\acrodef{OQAM}{offset quadrature amplitude modulation}
	\acrodef{OQPSK}{offset QPSK}
	\acrodef{OFDMA}{Orthogonal Frequency-division Multiple Access}
	\acrodef{OPEX}{Operating Expenditures}
	\acrodef{OQPSK/PM}{OQPSK with phase modulation}

	\acrodef{PAM}{pulse amplitude modulation}
	\acrodef{PAR}{peak-to-average ratio}
	\acrodef{PDF}{probability density function}
	\acrodef{PDP}{power dispersion profile}
	\acrodef{PMF}{probability mass function}                             
	\acrodef{PN}{pseudo-noise}
	\acrodef{PPM}{pulse position modulation}
	\acrodef{PRake}{Partial Rake}
	\acrodef{PSD}{power spectral density}
	\acrodef{PSEP}{pairwise synchronization error probability}
	\acrodef{PSK}{phase shift keying}
	\acrodef{PD}{power density}
	\acrodef{8-PSK}[$8$-PSK]{$8$-phase shift keying}
	\acrodef{PRS}{positioning reference signal}

	\acrodef{FSK}{frequency shift keying}

	\acrodef{QAM}{Quadrature Amplitude Modulation}
	\acrodef{QPSK}{quadrature phase shift keying}
	\acrodef{OQPSK/PM}{OQPSK with phase modulator }
	
	\acrodef{RD}[RD]{raw data}
	\acrodef{RDL}{"random data limit"}
	\acrodef{ric}[RIC]{restricted isometry constant}
	\acrodef{rict}[RICt]{restricted isometry constant threshold}
	\acrodef{rip}[RIP]{restricted isometry property}
	\acrodef{ROC}{receiver operating characteristic}
	\acrodef{rq}[RQ]{Raleigh quotient}
	\acrodef{RS}[RS]{Reed-Solomon}
	\acrodef{RSC}[RSSC]{RS based source coding}
	\acrodef{r.v.}{random variable}                               
	\acrodef{R.V.}{random vector}
	\acrodef{RMS}{root mean square}
	\acrodef{RFR}{radiofrequency radiation}
	\acrodef{RIS}{reconfigurable intelligent surface}
	\acrodef{RNA}{RiboNucleic Acid}
	
	\acrodef{SA}[SA-Music]{subspace-augmented MUSIC with OSMP}
	\acrodef{SCBSES}[SCBSES]{Source Compression Based Syndrome Encoding Scheme}
	\acrodef{SCM}{sample covariance matrix}
	\acrodef{SEP}{symbol error probability}
	\acrodef{SG}[SG]{sparse-land Gaussian model}
	\acrodef{SIMO}{single-input multiple-output}
	\acrodef{SINR}{signal-to-interference plus noise ratio}
	\acrodef{SIR}{signal-to-interference ratio}
	\acrodef{SISO}{single-input single-output}
	\acrodef{SMV}{single measurement vector}
	\acrodef{SNR}[\textrm{SNR}]{signal-to-noise ratio} 
	\acrodef{sp}[SP]{subspace pursuit}
	\acrodef{SS}{spread spectrum}
	\acrodef{SW}{sync word}
	\acrodef{SAR}{specific absorption rate}
	\acrodef{SSB}{synchronization signal block}

	\acrodef{TH}{time-hopping}
	\acrodef{ToA}{time-of-arrival}
	\acrodef{TR}{transmitted-reference}
	\acrodef{TW}{Tracy-Widom}
	\acrodef{TWDT}{TW Distribution Tail}
	\acrodef{TCM}{trellis coded modulation}
	\acrodef{TDD}{time-division duplexing}
	\acrodef{TDMA}{Time Division Multiple Access}
	
	\acrodef{UAV}{unmanned aerial vehicle}
	\acrodef{uric}[URIC]{upper restricted isometry constant}
	\acrodef{urict}[URICt]{upper restricted isometry constant threshold}
	\acrodef{UWB}{ultrawide band}
	\acrodef{UWBcap}[UWB]{Ultrawide band}   
	\acrodef{URLLC}{ultra reliable low latency communications}
	
	\acrodef{wuric}[UWRIC]{upper weak restricted isometry constant}
	\acrodef{wurict}[UWRICt]{UWRIC threshold}                
	\acrodef{UE}{user equipment}
	\acrodef{UL}{uplink}

	\acrodef{WiM}[WiM]{weigh-in-motion}
	\acrodef{WLAN}{wireless local area network}
	\acrodef{wm}[WM]{Wishart matrix}                               
	\acrodefplural{wm}[WM]{Wishart matrices}
	\acrodef{WMAN}{wireless metropolitan area network}
	\acrodef{WPAN}{wireless personal area network}
	\acrodef{wric}[WRIC]{weak restricted isometry constant}
	\acrodef{wrict}[WRICt]{weak restricted isometry constant thresholds}
	\acrodef{wrip}[WRIP]{weak restricted isometry property}
	\acrodef{WSN}{wireless sensor network}                        
	\acrodef{WSS}{wide-sense stationary}
	\acrodef{WHO}{World Health Organization}
	\acrodef{Wi-Fi}{wireless fidelity}
	
	\acrodef{sss}[SpaSoSEnc]{sparse source syndrome encoding}
	
	\acrodef{VLC}{visible light communication}
	\acrodef{VPN}{virtual private network} 
	\acrodef{RF}{radio frequency}
	\acrodef{FSO}{free space optics}
	\acrodef{IoST}{Internet of space things}
	
	\acrodef{GSM}{Global System for Mobile Communications}
	\acrodef{2G}{second-generation cellular network}
	\acrodef{3G}{third-generation cellular network}
	\acrodef{4G}{fourth-generation cellular network}
	\acrodef{5G}{5th-generation cellular network}	
	\acrodef{gNB}{next generation node B base station}
	\acrodef{NR}{New Radio}
	\acrodef{UMTS}{Universal Mobile Telecommunications Service}
	\acrodef{LTE}{Long Term Evolution}

	\acrodef{QoS}{quality of service}

	\acrodef{ULA}{uniform linear array}
	\acrodef{MS}{mobile station}
	\acrodef{6G}{sixth-generation cellular networks}
	\acrodef{EIV}{error-in-variable model}
	\acrodef{DoA}{direction of arrival}
	\acrodef{AoA}{angle of arrival}
	\acrodef{AoD}{angle of departure}
	\acrodef{LS}{least square}
	\acrodef{SLAM}{simultaneous localization and mapping}
	\acrodef{M-OMP}{multiple orthogonal matching pursuit}
	\acrodef{JLZA}{joint $\ell_{2,0}$ approximation}
	\acrodef{T-MSBL}{temporal multiple sparse Bayesian learning}
	\acrodef{MAP}{maximum a posteriori}
	\acrodef{ToA}{time of arrival}
	\acrodef{PSO}{particle swarm optimization}
	\acrodef{SDP}{semidefinite programming}
	\acrodef{ES}{exhaustive search}
	\acrodef{ISLAC}{integrated sensing, localization, and communication}
	\acrodef{SDS}{software defined surfaces}
	\acrodef{LiDAR}{light detection and ranging}
	\acrodef{3D}{three dimensional}
	\acrodef{ROI}{region of interest}
	\acrodef{DBSCAN}{density-based spatial clustering of applications with noise}
	\acrodef{2D}{two dimensional}
	\acrodef{SMRF}{simple morphological filter}
	\acrodef{EM}{electromagnetic}
    \acrodef{MLE}{maximum likelihood estimator}
    \acrodef{PEB}{position error bound}
    \acrodef{FIM}{Fisher information matrix}

\begin{document}
\def\tr{{\mathrm{tr}}}

\def\It{I_{\mathsf{t}}}
\def\T{\emph{T}}
\def\R{\emph{R}}
\def\S{\emph{S}}
\def\t{\mathsf{t}}
\def\r{\mathsf{r}}
\def\s{\mathsf{s}}
\def\rt{{\boldsymbol{\rho}}_{\mathsf{t}}}
\def\rs{{\boldsymbol{\rho}}_{\mathsf{s}}}
\def\rr{{\boldsymbol{\rho}}}
\def\ro{\boldsymbol{\rho}}
\def\at{x_{\mathsf{t}}}
\def\bt{y_{\mathsf{t}}}
\def\ct{z_{\mathsf{t}}}
\def\ar{x_{\mathsf{r}}}
\def\br{y_{\mathsf{r}}}
\def\ccr{z_{\mathsf{r}}}
\def\cs{z_{\mathsf{s}}}
\def\as{x_{\mathsf{s}}}
\def\bs{y_{\mathsf{s}}}
\def\ko{k_{0}}
\def\kd{k_\mathsf{d}}
\def\epsd{\epsilon_{\textrm{r}}}
\def\epso{\epsilon_{\textrm{0}}}
\def\muo{\mu_{\textrm{0}}}

\def\vst{\boldsymbol{\rho}_{\mathsf{s}\mathsf{t}}}
\def\vrs{\boldsymbol{\rho}_{\mathsf{r}\mathsf{s}}}

\def\rb{{\rho}_{b}}

\newcommand{\rts} {\rho_{\mathsf{ts}}}
\newcommand{\rsr} {\rho_{\mathsf{sr}}}
\newcommand{\rtr} {\rho_{\mathsf{tr}}}
\newcommand{\rtrb} {\rho_{\mathsf{tr}b}}
\newcommand{\rot} {\rho_{\mathsf{ot}}}

\newcommand{\Ho}[1] { H_{0}^{(2)}\left({#1}\right) }
\newcommand{\Hl}[1] { H_{1}^{(2)}\left({#1}\right) }
\newcommand{\Hol}[1] { H_{0}^{(1)}\left({#1}\right) }
\newcommand{\Hll}[1] { H_{1}^{(1)}\left({#1}\right) }
\newcommand{\Jo} {J_{0}}
\newcommand{\Jl} {J_{1}}
\newcommand{\Yo} {Y_{0}}
\newcommand{\Yl} {Y_{1}}

\newcommand{\Nb} {N_\mathsf{B}}
\newcommand{\Nm} {N_\mathsf{M}}
\newcommand{\Nl} {N_\mathsf{R}}
\newcommand{\Ns} {N}
\newcommand{\Hm} {\mathbf{H}}
\newcommand{\y} {\mathbf{y}}
\newcommand{\x} {\mathbf{x}}
\newcommand{\n} {\boldsymbol{\omega}}
\newcommand{\Hbm} {\mathbf{H}_\mathsf{BM}}
\newcommand{\Hbl} {\mathbf{H}_\mathsf{BR}}
\newcommand{\Hlm} {\mathbf{H}_\mathsf{RM}}
\newcommand{\bmOmega} {\bm{\Omega}}
\newcommand{\alphar} {\bm{\alpha}_\mathsf{r}}
\newcommand{\alphat} {\bm{\alpha}_\mathsf{t}}
\newcommand{\rhobm} {\rho_\mathsf{BM}}
\newcommand{\rhobl} {\rho_\mathsf{BR}}
\newcommand{\rholm} {\rho_\mathsf{RM}}
\newcommand{\rhoblm} {\rho_\mathsf{BRM}}

\newcommand{\tbm} {\tau_{bm}}
\newcommand{\tbl} {\tau_{br}}
\newcommand{\tlm} {\tau_{rm}}
\newcommand{\tdm} {\tau_{sm}}

\newcommand{\taubm} {\tau_\mathsf{BM}}
\newcommand{\taubl} {\tau_\mathsf{BR}}
\newcommand{\taulm} {\tau_\mathsf{RM}}
\newcommand{\taudm} {\tau_\mathsf{DM}}

\newcommand{\dbmo} {d_\mathsf{BM}}
\newcommand{\ddmo} {d_\mathsf{SM}}
\newcommand{\dblo} {d_\mathsf{BR}}
\newcommand{\dlmo} {d_\mathsf{RM}}

\newcommand{\xilm} {\xi_{\mathsf{RM}}}
\newcommand{\etal} {\eta_r}
\newcommand{\etam} {\eta_m}

\newcommand{\xibm} {\xi_{\mathsf{BM}}}

\newcommand{\chibm}{\chi_{\mathsf{B}\mathsf{M}}}
\newcommand{\chiblm}{\chi_{\mathsf{B}\mathsf{R}\mathsf{M}}}

\newcommand{\phibm} {\phi_\mathsf{BM}}
\newcommand{\phibl} {\phi_\mathsf{BR}}
\newcommand{\philm} {\phi_\mathsf{RM}}
\newcommand{\phiL} {\phi_{\mathsf{R}}}
\newcommand{\phidm} {\phi_\mathsf{SM}}

\newcommand{\thetadm} {\theta_\mathsf{SM}}
\newcommand{\thetabm} {\theta_\mathsf{BM}}
\newcommand{\thetabl} {\theta_\mathsf{BR}}
\newcommand{\thetalm} {\theta_\mathsf{RM}}

\newcommand{\thetam} {\theta_m}
\newcommand{\thetab} {\theta_b}
\newcommand{\thetal} {\theta_l}

\newcommand{\xms} {x_{\mathsf{M}}}
\newcommand{\yms} {y_{\mathsf{M}}}
\newcommand{\zms} {z_{\mathsf{M}}}

\newcommand{\phim} {\phi_m}
\newcommand{\phib} {\phi_b}
\newcommand{\phil} {\phi_r}

\newcommand{\thetaL} {\theta_{\mathsf{R}}}

\newcommand{\dm} {d_{m}}
\newcommand{\dl} {d_{r}}
\newcommand{\db} {d_{b}}
\newcommand{\dbm} {d_{bm}}
\newcommand{\ddm} {d_{sm}}
\newcommand{\dbl} {d_{br}}
\newcommand{\dlm} {d_{rm}}
\newcommand{\fn} {f_n}

\newcommand{\xb}{x_b}
\newcommand{\yb}{y_b}
\newcommand{\nb}{w_b}
\newcommand{\zb}{z_b}

\newcommand{\xm}{x_m}
\newcommand{\ym}{y_m}
\newcommand{\zm}{z_m}

\newcommand{\xd}{x_s}
\newcommand{\yd}{y_s}
\newcommand{\zd}{z_s}

\newcommand{\Cbrm}{C_{brm}}
\newcommand{\BCr}{\bar{C}_{r}}
\newcommand{\BCk}{\bar{C}_{k}}
\newcommand{\Cbkm}{C_{bkm}}
\newcommand{\betam}{\beta_{m}}
\newcommand{\fc}{f_\mathsf{c}}
\newcommand{\thetar}{\theta_r}
\newcommand{\thetak}{\theta_k}

\newcommand{\xdi}{x_d^{(0)}}
\newcommand{\ydi}{y_d^{(0)}}
\newcommand{\zdi}{z_d^{(0)}}

\newcommand{\xmi}{x_m^{(0)}}
\newcommand{\ymi}{y_m^{(0)}}
\newcommand{\zmi}{z_m^{(0)}}

\newcommand{\cosam}{c_{\alpha_{\mathsf{M}}}}
\newcommand{\sinam}{s_{\alpha_{\mathsf{M}}}}
\newcommand{\cosbm}{c_{\beta_{\mathsf{M}}}}
\newcommand{\sinbm}{s_{\beta_{\mathsf{M}}}}

\newcommand{\cosgm}{c_{\gamma_{\mathsf{M}}}}
\newcommand{\singm}{s_{\gamma_{\mathsf{M}}}}

\newcommand{\xl}{x_r}
\newcommand{\yl}{y_r}
\newcommand{\zl}{z_r}
\newcommand{\godm}{\mathsf{G}^{(1)}_{sm}}
\newcommand{\gobm}{\mathsf{G}^{(1)}_{bm}}
\newcommand{\gobl}{\mathsf{G}^{(1)}_{bl}}
\newcommand{\gtbm}{\mathsf{G}^{(2)}_{bm}}
\newcommand{\gtdm}{\mathsf{G}^{(2)}_{sm}}
\newcommand{\gtBM}{G_{\mathsf{B,M}}^{(b,m)}}
\newcommand{\gtDM}{G_{\mathsf{D,M}}^{(d,m)}}

\newcommand{\gtbl}{\mathsf{G}^{(2)}_{br}}
\newcommand{\gtlm}{\mathsf{G}^{(2)}_{rm}}
\newcommand{\gtLM}{G_{\mathsf{R,M}}^{(r,m)}}

\newcommand{\gtBL}{G_{\mathsf{B,R}}^{(b,r)}}

\newcommand{\xmo}{x_{\mathsf{M}}}
\newcommand{\ymo}{y_{\mathsf{M}}} 
\newcommand{\zmo}{z_{\mathsf{M}}} 
\newcommand{\pmo}{\mathbf{p}_{\mathsf{M}}}

\newcommand{\pdo}{\mathbf{p}_{\mathsf{D}}}
\newcommand{\ado}{\boldsymbol{\Phi}_{\mathsf{D}}}
\newcommand{\amo}{\boldsymbol{\phi}_{\mathsf{M}}}
\newcommand{\abo}{\boldsymbol{\Phi}_{\mathsf{B}}}
\newcommand{\alo}{\boldsymbol{\Phi}_{\mathsf{R}}}

\newcommand{\am}{{a}_{\mathsf{M}}}

\newcommand{\xbo}{x_{\mathsf{B}}}
\newcommand{\ybo}{y_{\mathsf{B}}} 
\newcommand{\zbo}{z_{\mathsf{B}}} 
\newcommand{\pbo}{\mathbf{p}_{\mathsf{B}}}

\newcommand{\xlo}{x_{\mathsf{R}}}
\newcommand{\ylo}{y_{\mathsf{R}}} 
\newcommand{\zlo}{z_{\mathsf{R}}} 
\newcommand{\plo}{\mathbf{p}_{\mathsf{R}}}

\newcommand{\hbm}{h_{bm}}
\newcommand{\hbl}{h_{br}}
\newcommand{\hlm}{h_{rm}}
\newcommand{\omegal}{\exp\left({j\,\thetar}\right)}
\newcommand{\xbs}{{x}_{\mathsf{B}}}
\newcommand{\ybs}{{y}_{\mathsf{B}}}
\newcommand{\zbs}{{z}_{\mathsf{B}}}
\newcommand{\xds}{{x}_{\mathsf{D}}}
\newcommand{\yds}{{y}_{\mathsf{D}}}
\newcommand{\zds}{{z}_{\mathsf{D}}}
\newcommand{\xls}{{x}_{\mathsf{R}}}
\newcommand{\yls}{{y}_{\mathsf{R}}}
\newcommand{\zls}{{z}_{\mathsf{R}}}
\newcommand{\dant}{{d}_{\mathsf{ant}}}
\newcommand{\muBM} {\mu_{b,\mathsf{BM}}[n]}
\newcommand{\muBLM} {\mu_{b,\mathsf{BRM}}[n]}
\newcommand{\muBMc} {\mu^{*}_\mathsf{BM}[n]}
\newcommand{\muBLMc} {\mu^{*}_\mathsf{BRM}[n]}

\newcommand{\mubm} {\mu_{bm}[n]}
\newcommand{\cmubmp} {\mu_\mathsf{bm'}^*[n]}

\newcommand{\mublm} {\mu_{brm}[n]}
\newcommand{\mubmc} {\mu^{*}_\mathsf{bm}[n]}
\newcommand{\mublmc} {\mu^{*}_\mathsf{brm}[n]}

\newcommand{\alphamo}{\alpha_{\mathsf{M}}}
\newcommand{\betamo}{\beta_{\mathsf{M}}}
\newcommand{\gammamo}{\gamma_{\mathsf{M}}}

\newcommand{\Phim}{\Phi_{\mathsf M}}

\newcommand{\PEB}{\mathsf{PEB}}
\newcommand{\OEB}{\mathsf{OEB}}

\title{Electromagnetic Channel Model and Sensing Performance for ISAC: The Case of a Single Cylindrical Scatterer}

\author{Gabriele~Gradoni,~\IEEEmembership{Member,~IEEE,} Ahmed~Elzanaty,~\IEEEmembership{Senior Member,~IEEE,}
Yadong~Yang,
Mohsen~Khalily,~\IEEEmembership{Senior Member,~IEEE,} 
Mahmoud~Shahabadi,~\IEEEmembership{Senior Member,~IEEE,} 
Fan~Wang,
Peiying~Zhu,~\IEEEmembership{Fellow,~IEEE,}
Wen~Tong,~\IEEEmembership{Fellow,~IEEE}
and~Rahim~Tafazolli,~\IEEEmembership{Fellow,~IEEE}
\thanks{G. Gradoni, A. Elzanaty, Y. Yang, M. Khalily, M. Shahabadi, R. Tafazolli are with the 6GIC, Institute for Communication Systems (ICS), University of Surrey, Guildford, GU2 7XH, United Kingdom (e-mail: \{g.gradoni, a.elzanaty, 
yadong.yang,
m.khalily, m.shahabadi, r.tafazolli\}@surrey.ac.uk).}
\thanks{Fan Wang, Peiying Zhu, and Wen Tong are with Huawei Technologies Canada Co., Ltd., Ottawa, ON K2K 3J1, Canada.}
}

\maketitle

\begin{abstract}
We propose an electromagnetic wireless channel model for accurate target localization in radar-like sensing scenarios. 
    The model employs a self-consistent method grounded in wave scattering theory and imposing boundary conditions, providing the (impedance) channel transfer matrix for a propagation channel with a single target  modeled as a cylindrical scatterer.
The model is: 1) physics-based, as it includes electromagnetic interactions between objects; 2) unified, as it is valid in both near- and far-field regions. Within the canonical two-dimensional and 2.5-dimensional configurations with line-source and dipole-source excitations, respectively, the model supports dielectric cylinders with frequency-dependent permittivity.
Localization accuracy is assessed via the  Cramér–Rao lower bound for a dielectric cylinder with variable position, radius, and dielectric constant. 
Provisions for future extensions to extended targets are discussed in the context of 3GPP channel modeling.
\end{abstract}

\IEEEpeerreviewmaketitle

\section{Introduction}

\IEEEPARstart{I}{ntegrated} sensing and communications (ISAC), also known as joint sensing and communications (JSAC), aims to incorporate sensing capabilities into mobile wireless networks. This includes localizing active users and passive objects by leveraging the widespread deployment of base stations (BSs) across large geographical areas. Network architectures that realize these functionalities are often referred to as communication-centric ISAC/JSAC systems~\cite{liu2020joint,MasourosTWC:18}.

The ability of electromagnetic (EM) waves to penetrate walls and propagate within buildings enables the exploitation of non-line-of-sight (NLOS) energy for applications such as behind-the-wall imaging. 
Hence, it is crucial to characterize the spatial consistency of the wireless channel in the presence of antenna arrays, designated targets, and environmental objects \cite{ElzanatyAnnaFrans:21}. 
This requires a model that accounts for pervasive wave scattering and captures the multiple EM interactions between all channel components in a self-consistent manner~\cite{ElzanatyMismatch:24,AlexElzanaty:25}.
 
More specifically, from a sensing and localization perspective, achievable positioning accuracy is fundamentally determined by the underlying EM field observations. Performance limits such as the Cramér–Rao lower bound (CRLB) depend explicitly on the amplitude and phase of the received fields, as well as on how these quantities vary with respect to the target parameters. While radar-oriented metrics such as the radar cross section (RCS) are widely used to characterize scattering strength, they are inherently quadratic, power-based quantities and therefore discard phase information. As a result, RCS-based descriptions, whether in far-field or near-field formulations, are inaccurate on their own for estimating the position of the target or deriving insightful localization bounds.

    Current literature on ISAC channel modeling offers three main approaches: deterministic~\cite{luo2024channel}, statistical~\cite{zhang2024cluster}, and hybrid~\cite{zhang2026channel}.  Statistical models offer scalability  but abstract away the physical relationship between the EM fields and the target properties, which is critical for localization.
    For deterministic channel modeling, conventional ray-tracing (RT) techniques are widely adopted for EM propagation analysis in large-scale environments. However, RT tools such as those presented in \cite{fugen2006capability} and \cite{Reichardt2011} are fundamentally based on geometric optics approximations, where EM waves are represented as discrete rays undergoing reflection, transmission, and diffraction along predefined propagation paths. As a consequence, many wave phenomena are only approximately modeled or entirely neglected, including near-field interactions, polarization coupling, and phase-coherent scattering effects. These limitations become particularly critical in ISAC scenarios, where sensing performance often relies on accurate phase-resolved scattering responses from targets and surrounding structures. Furthermore, RT simulations become increasingly computationally demanding as the environmental complexity grows. Accurately resolving propagation in dense urban scenes often requires launching millions of rays and handling numerous reflections and diffractions, which significantly increases computational overhead and memory usage. Recent developments, such as GPU-accelerated ray engines (e.g., NVIDIA OptiX), have improved the efficiency of ray traversal and intersection calculations~\cite{Zhang2021}. Nevertheless, these implementations primarily accelerate geometric ray processing and do not address the intrinsic limitations of ray-based propagation models in representing full electromagnetic interactions.

Beyond RT, several physics-based EM channel models have been proposed to preserve full-wave consistency while remaining analytically tractable~\cite{miller2000communicating,dardari2020communicating,pizzo2023wide,pizzo2022spatial}. Representative examples in the literature include:
(i) EM mode-based models for wave coupling between finite regions, which characterize propagation via eigenmode decompositions of the electromagnetic operator between volumes or surfaces \cite{miller2000communicating,dardari2020communicating}. These models enable rigorous analysis of spatial degrees-of-freedom (DOF) and coupling strengths under line-of-sight (LOS) propagation, but do not incorporate NLOS scattering mechanisms arising from scatterers in the environment;
    (ii) reflection-based models for wide-aperture MIMO, which describe propagation via plane-wave spectra and Fresnel/image-theory reflections from flat, smooth, infinitely extended planar dielectric interfaces~\cite{pizzo2023wide}. These models enable rigorous analysis of DOF/capacity, but rely on an idealized infinitely extended planar interface and do not provide an explicit finite-sized target model whose position/size/material parameters enter the channel in a differentiable form suitable for sensing/localization; and
(iii) more general plane-wave/scattering-operator formulations, which model arbitrary NLOS propagation as a linear mapping between transmitted and received plane-wave spectra, capturing angular coupling as well as near-/far-field effects, yet typically parameterize scattering at the level of angular kernels and/or (possibly) random-field descriptions rather than explicit, differentiable target geometry/material parameters~\cite{pizzo2022spatial}. Consequently, while these approaches are well suited for communication-centric channel characterization (e.g., dimensionality/DoF), they are not directly applicable to ISAC sensing/localization, where explicit channel dependence on target parameters (e.g., position, size, permittivity) is required for target-specific estimation.

These observations highlight a key gap between these communication-centric  channel models and ISAC sensing requirements. In ISAC, the objects of interest are not merely part of a fixed environment but constitute unknown targets whose location and physical properties must be inferred from the received fields. This calls for channel models that provide an explicit and differentiable mapping from target parameters (e.g., position, size, and material constants) to the complex-valued channel response, so that target-specific information can be physically embedded into the propagation model and subsequently exploited for localization and parameter estimation.

        To meet these needs, we develop a physics-based wireless channel model derived from first EM principles, with the goal of enabling target-parameter–dependent sensing analysis. A necessary step is to solve a single-target scattering problem that provides an explicit and differentiable mapping from target parameters to the channel response, which can later be embedded into multi-scattering. To obtain closed-form, phase-resolved expressions that remain amenable to parameter-sensitivity analysis, we adopt a controlled canonical configuration: we consider time-harmonic excitation and model the target as an infinitely long  cylinder. Within this canonical geometry, the choice of the illumination becomes an integral part of the sensing model, since different source types induce distinct field distributions and phase structures around the scatterer, thereby directly impacting the resulting  achievable localization performance bounds~\cite{pizzo2022fourier}. Accordingly, the cylinder is illuminated by either an infinitely extended thin current line or a localized dipole source. Within this analytically tractable formulation, we derive closed-form expressions for the scattered fields and an electromagnetic definition of the channel transfer matrix tailored to single-target localization in ISAC following electromagnetic theory~\cite{chew2001fast,chew1993huygens}. Starting from the general scattering solution, a dipole approximation is further derived to yield simplified expressions for electrically small dielectric cylinders~\cite{he2020cascaded}, which constitute a canonical yet challenging scenario for precise localization.

Unlike approaches that rely on numerical meshing or volumetric discretization such as the finite-difference time-domain (FDTD)\cite{yee14numerical} and finite-element method (FEM)\cite{jin2015finite}, our formulation leverages closed-form analytical expressions, so that the number of unknowns scales primarily with the number of scattering objects rather than the size of the computational domain. This leads to substantial efficiency gains and facilitates scalable analysis when combined with higher-level frameworks such as stochastic geometry. More importantly for sensing, the proposed model explicitly computes scattering coefficients by enforcing boundary conditions on non-flat target surfaces. This contrasts with common reflection-coefficient models that are tailored to plane-wave incidence on flat interfaces~\cite{fugen2006capability}. 

Furthermore, we employ the rigorous scattering model to quantify the CRLB in closed-form mathematical expressions. Numerical results are provided for dielectric cylinders, across various positions, radii, and dielectric constants. The theoretical channel transfer matrix is validated using commercial full-wave EM simulation software. The results are of significant interest for the design and performance evaluation of next-generation ISAC technologies.

\subsection{Novelty of the Paper} 
We introduce a mathematical model that captures the full-wave electromagnetic propagation of wireless signals through complex environments containing scattering objects. Specifically, we develop fundamental aspects of the channel model tailored for communication-centric ISAC systems, addressing the following key components:
\begin{enumerate}
    \item Derivation from the  electromagnetic theory enforcing boundary conditions on a canonical cylindrical target;
    \item A unified cylindrical-wave scattering formulation, beyond idealized plane-wave excitation, in which near-field and far-field behaviors arise from the same underlying representation;

    \item Angular dispersion arising from varying incidence and scattering angles
through the cylindrical-wave scattering expansion;
    \item Frequency dependence and material dispersion captured explicitly via the frequency-dependent wavenumbers and the cylinder’s material parameters in the scattering coefficients;
\item {A direct connection between the complex scattered field and target-localization performance, where both amplitude and phase are retained to evaluate the position error bound}.    
\end{enumerate}
Importantly, the model provides a field-based formulation that asymptotically retrieves the canonical RCS when the target lies in the near/far field of both the transmitter and receiver arrays. Crucially, the phase information embedded in the scattered fields is preserved, which is essential for accurate localization and sensing of dielectric targets.

\subsection{Notation} 
We denote matrices and vectors using bold capital and lowercase letters, respectively; their individual elements are represented in plain text with subscripts. The operators $(\cdot)^{T}$, $(\cdot)^{H}$, and $(\cdot)^{*}$ denote the transpose, Hermitian (conjugate transpose), and complex conjugate, respectively. The norm $\| \cdot \|$ indicates the L$2$ norm for vectors and the spectral norm for matrices, while $\| \cdot \|_{\mathrm{F}}$ denotes the Frobenius norm. The trace and diagonal operators are written as $\mathrm{tr}(\cdot)$ and $\mathrm{diag}(\cdot)$, respectively. The gradient with respect to the $N \times 1$ vector $\mathbf{x}$ is given by $\nabla_{\mathbf{x}} = [\frac{\partial}{\partial x_1}, \dots, \frac{\partial}{\partial x_N}]^{T}$.  The Hankel function of the $k$-th type and the $n$-th order is denoted by $H_{n}^{(k)}(\cdot)$. 

\subsection{Outline of the Paper} 
Section~\ref{Sec:Summary} provides an overview of the electromagnetic channel model along with the main analytical results. Section~\ref{sec:Theoretical Framework} details the theoretical development of the scattering model, including formulations for both line source excitation (2D model) and dipole excitation (2.5D model). Section~\ref{Sec:Sensing} presents the sensing capabilities of the proposed models, focusing on localization performance and the derivation of the CRLB. Numerical results are discussed in Section~\ref{Sec:NumericalResults}, including a parametric study on a single dielectric cylinder validated through full-wave electromagnetic simulations and evaluation of localization algorithms. Finally, Sections~\ref{sec:Challenges} and~\ref{sec:conclusion} conclude the paper by outlining open challenges and summarizing key insights.

\section{Summary of the Results}
\label{Sec:Summary}
{
\begin{theorem}[Received field with line source excitation in the presence of a cylinder] \label{th:line}
Let $\T$ be an electric current
filament with a time-harmonic variation of $e^{j \omega t}$ and
a phasor of $\It$, located at $\rt = [\at,\bt] = \at \hat{\bf x} + \bt \hat{\bf y} $ where $\hat{\bf x}$ and $\hat{\bf y}$ are the unit vectors in the directions of the $\emph{x}$ and $\emph{y}$ axes, respectively. Let $\R$ be a receiver located at $\rr = [\ar , \br]$.  Let a scatterer, i.e., the target, $\S$ be present in the environment and located at $\rs = [\as , \bs]$. The scatterer is assumed to
be an infinite dielectric cylinder of radius $a$ whose relative permittivity in the frequency band of interest is known to be a complex function of frequency $\epsd(\omega)$. 
Let ${\rho}_{\mathsf{{e}_{1}}\mathsf{{e}_{2}}} =\|\boldsymbol{\rho}_{\mathsf{{e}_{1}}}-\boldsymbol{\rho}_{\mathsf{{e}_{2}}}\|$ where $\mathsf{{e}_{1}}$, $\mathsf{{e}_{2}} \in  \{ \r,\t,\s \}$ be the distance between two entities $\mathsf{{e}_{1}}$ and $\mathsf{{e}_{2}}.$ 
Let $\epsd$, $\omega$, $\ko\triangleq \omega \sqrt{\muo \epso}$, $\kd\triangleq \omega \sqrt{\muo \epso \epsd}$  denote the permittivity of the dielectric scatterer, the angular operating frequency, the wave number in free space, and the wave number in the dielectric cylinder, respectively.
Define the receiver and transmitter azimuths relative to the cylinder center as $\phi_{\rm r}\triangleq\operatorname{atan2}(\br-\bs,\ar-\as)$ and $\phi_{\rm t}\triangleq\operatorname{atan2}(\bt-\bs,\at-\as)$, respectively. Using a finite number of cylindrical modes, the electric field at the receiver is approximated as
\begin{equation}\label{eqn:Ez_line}
\begin{split}
 E_z(\ar,\br)\simeq{}&-\frac{\omega\muo}{4}\Bigg[\It H_0^{(2)}(\ko\rtr)\\
 &+\sum_{n=-N}^{N}I_n^{\textrm{s}}H_n^{(2)}(\ko\rsr)
 e^{jn(\phi_{\rm r}-\phi_{\rm t})}\Bigg],
\end{split}
\end{equation}
where, for every retained modal order $n\in\{-N,\ldots,N\}$,
\begin{equation}\label{eqn:summary_line_modal_coefficient}
I_n^{\textrm{s}}=\Gamma_{d,n}(\omega,a,\epsd)
H_n^{(2)}(\ko\rts)\It,
\end{equation}
with
\begin{equation}\label{eqn:summary_line_scattering_coefficient}
\Gamma_{d,n}=\frac{\dfrac{\kd}{\ko}J_n'(\kd a)J_n(\ko a)-J_n(\kd a)J_n'(\ko a)}{J_n(\kd a)H_n^{(2)'}(\ko a)-\dfrac{\kd}{\ko}J_n'(\kd a)H_n^{(2)}(\ko a)}.
\end{equation}
Here, $N$ is the required modal truncation order, which depends primarily on the electrical size of the cylinder, characterized by $|\kd|a$. Larger electrical sizes generally require higher-order modes.
\end{theorem}

\begin{proof}
Applying the tangential-field boundary conditions at $\rho=a$ and eliminating the internal-field coefficient gives \eqref{eqn:summary_line_modal_coefficient}--\eqref{eqn:summary_line_scattering_coefficient}; see Section~\ref{subsec:lineexcitation}.
\end{proof}

\begin{corollary}[Electrically small cylinder under line source excitation]\label{cor:line_n0} Under the setting of Theorem II.1, suppose that the cylinder is electrically small, $|k_d|a \ll 1$, so that the zeroth-order modal scattering coefficient dominates the scattered field. Then
\begin{equation}\label{eqn:Ez_line_n0}
E_z(\ar,\br)\simeq-\frac{\omega\muo}{4}
\left[\It H_0^{(2)}(\ko\rtr)+I_0^{\textrm{s}}H_0^{(2)}(\ko\rsr)\right].
\end{equation}
If, additionally, $\ko\rsr\gg1$ and $\ko\rtr\gg1$, \eqref{eqn:Ez_line_n0} admits the large-distance asymptotic form
\begin{equation}\label{eqn:Ez_line_far}
\begin{split}
E_z(\ar,\br)\sim{}&-\frac{\omega\muo}{4}
\sqrt{\frac{2}{\pi\ko}}e^{j\pi/4}\\
&\times\left[
\frac{\It e^{-j\ko\rtr}}{\sqrt{\rtr}}
+\frac{I_0^{\textrm{s}}e^{-j\ko\rsr}}{\sqrt{\rsr}}
\right].
\end{split}
\end{equation}
\end{corollary}

\begin{proof}
Retaining only the $n=0$ term in \eqref{eqn:Ez_line} and using $J_0'(x)=-J_1(x)$ and $H_0^{(2)'}(x)=-H_1^{(2)}(x)$ in \eqref{eqn:summary_line_scattering_coefficient} gives \eqref{eqn:Ez_line_n0}; see Section~\ref{subsec:lineexcitation}. Under the additional large-distance conditions, using
\begin{equation}\label{eqn:H2_large}
H_0^{(2)}(z)\sim\sqrt{\frac{2}{\pi z}}e^{-j(z-\pi/4)},
\qquad z\to\infty,
\end{equation}
for the two Hankel functions explicitly appearing in \eqref{eqn:Ez_line_n0} yields \eqref{eqn:Ez_line_far}.
\end{proof}
\begin{figure}[t!]
\centering
\includegraphics[scale=0.08]{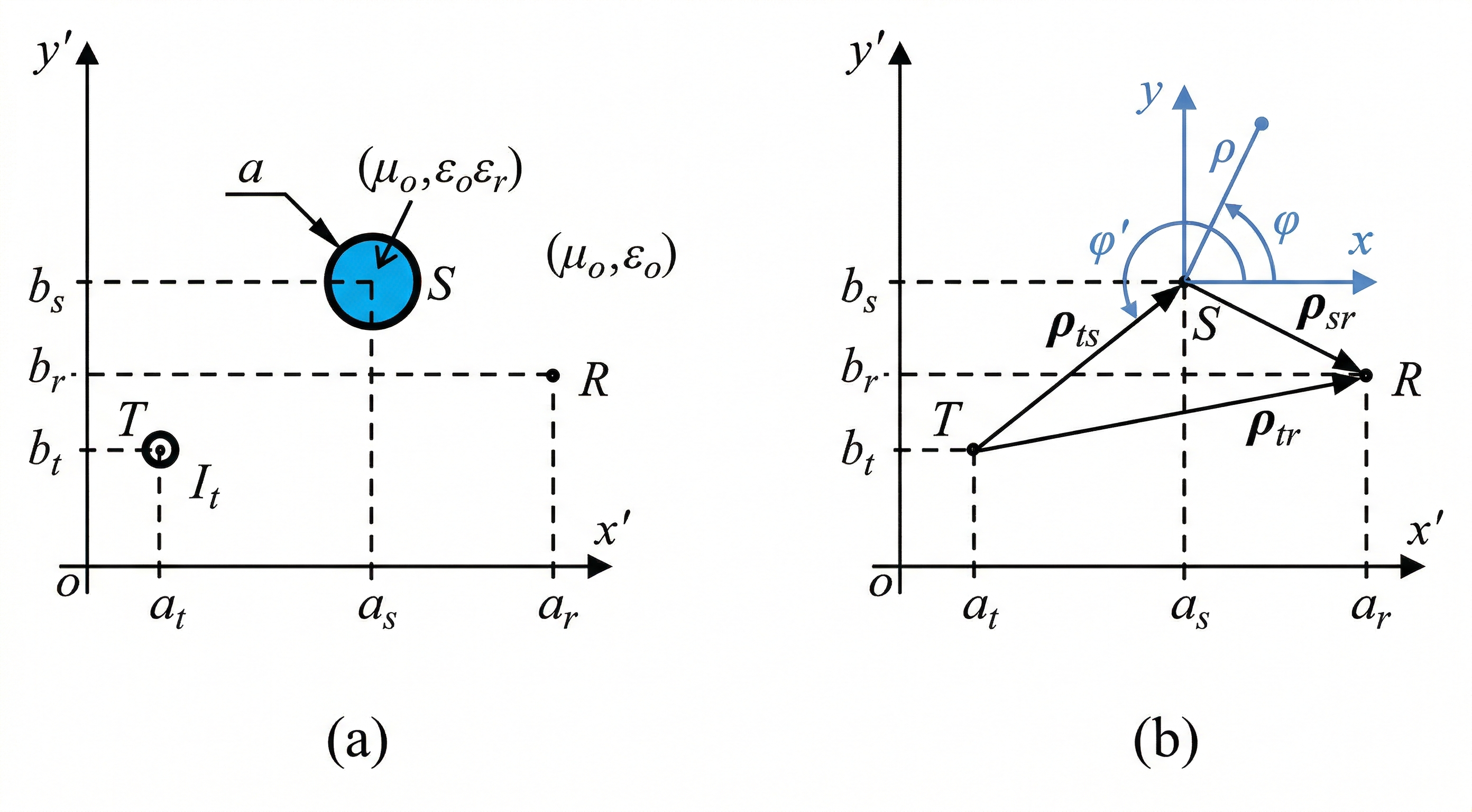}
\caption{The scenario represented here includes (a) the configuration of the two-dimensional scattering channel; (b) the coordinate system and the relative position vectors required for the proposed formulation.}
\label{fig:Config}
\end{figure}
\begin{corollary}[Perfectly Conducting Cylinder]
For a perfectly conducting cylinder under line-source excitation, the complete model in Theorem~\ref{th:line} remains valid after replacing the dielectric modal coefficients by
\begin{equation}
\Gamma_{c,n}(\omega,a)=-\frac{J_n(\ko a)}{H_n^{(2)}(\ko a)},\qquad n\in\mathbb Z,
\end{equation}
where $I_n^{\textrm{s}}=\Gamma_{c,n}H_n^{(2)}(\ko\rts)\It$. Its electrically small, zeroth-order corollary is obtained with $\Gamma_{c,0}=-J_0(\ko a)/H_0^{(2)}(\ko a)$.
\end{corollary}

\begin{theorem}[Received field with dipole excitation in the presence of a cylinder scatterer]\label{th:dipole}

Let $\T$ be an infinitesimal dipole transmit antenna located at
$\rt = [\at,\bt,0] = \at \hat{\bf x} + \bt \hat{\bf y}+0 \hat{\bf z}$,
where $\hat{\bf x}$, $\hat{\bf y}$, and $\hat{\bf z}$ are the unit
vectors along the $x$, $y$, and $z$ axes, respectively.
Let $\R$ be a receiver located at
$\boldsymbol{\rho}_{\rm r} = [\ar,\br,0]$.
Let a scatterer, that is, the target, $\S$, be present in the
environment. The scatterer is assumed to be an infinitely long
dielectric cylinder of radius $a$, oriented along the $z$-axis,
whose axis intersects the $z=0$ plane at
$\rs = [\as,\bs,0]$.
Its relative permittivity in the frequency band of interest
is known to be a complex function of frequency $\epsd(\omega)$.

Let
$\boldsymbol{\rho}_{\mathsf{e}_1\mathsf{e}_2}
=\boldsymbol{\rho}_{\mathsf{e}_1}
-\boldsymbol{\rho}_{\mathsf{e}_2}$
be the vector between two entities $\mathsf{e}_1$ and
$\mathsf{e}_2$, where
$\mathsf{e}_1,\mathsf{e}_2\in\{\r,\t,\s\}$.
Let
$\rho_{\mathsf{e}_1\mathsf{e}_2}
=\|\boldsymbol{\rho}_{\mathsf{e}_1}
-\boldsymbol{\rho}_{\mathsf{e}_2}\|$
denote their transverse distance on the $z=0$ plane.

Let $\phi_{\rm r}$ and $\phi_{\rm t}$ denote the receiver
and transmitter azimuths relative to the cylinder axis.
The received electric field is
\begin{equation}
   E_z = E^{\text{i}}_{z} + E^{\text{s}}_{z}
   \label{eq:EVDipole}
\end{equation}
where
\begin{align}
 E^{i}_z
 &= - j \omega \muo I^t dl
 \frac{e^{-j \ko \rtr}}{4 \pi \rtr}
 \left(
 1 - j \frac{1}{\ko \rtr}
 -\frac{1}{\left(\ko \rtr\right)^2}
 \right),\\
 E^{\text{s}}_{z}
 &\simeq-\frac{\omega\muo I^t dl}{8\pi}
 \sum_{n=-N}^{N}e^{jn(\phi_{\rm r}-\phi_{\rm t})}
 \int_{-\infty}^{\infty}
 \left(1-\frac{k_z^2}{\ko^2}\right)
 \Gamma_n(k_z)\nonumber\\
 &\quad\times H_n^{(2)}(k_\rho\rts)
 H_n^{(2)}(k_\rho\rsr)\,dk_z.
 \label{eqn:summary_dipole_full}
\end{align}
Here $k_\rho^2+k_z^2=\ko^2$ and
\begin{equation}
\Gamma_n(k_z)\triangleq
[\mathbf A_n^{-1}(k_z)\mathbf c_n(k_z)]_1,
\end{equation}
where $\mathbf A_n$ and $\mathbf c_n$ are defined in
\eqref{eqn:FieldCont}. As in the line-source model,
the required $N$ grows with the electrical size,
principally $|\kd|a$.

The scattered field is evaluated at $z=0$ to obtain
the total field on the same plane as the incident field.

\end{theorem}

\begin{proof}
Please refer to Section~\ref{subsec:dipole}.
\end{proof}

\begin{corollary}[Electrically small cylinder under dipole excitation]\label{cor:dipole_n0}
Under the setting of Theorem~\ref{th:dipole}, if the cylinder is electrically small, i.e., $|\kd|a\ll 1$, the zeroth-order ($n=0$) contribution dominates the scattered field. Then
\begin{multline}\label{eqn:summary_dipole_n0}
E_z\simeq E_z^{\rm i}-\frac{\omega\muo I^t dl}{8\pi}
\int_{-\infty}^{\infty}\left(1-\frac{k_z^2}{\ko^2}\right)\Gamma_0(k_z)\\
\times H_0^{(2)}(k_\rho\rts)H_0^{(2)}(k_\rho\rsr)\,dk_z.
\end{multline}
\end{corollary}

\begin{proof}
The result follows by retaining the zeroth-order term of the converged modal summation in \eqref{eqn:summary_dipole_full}.
\end{proof}

\begin{proof}
If the scatterer is a perfectly conducting cylinder, the two boundary
conditions for the dielectric cylinder are reduced to a single boundary
condition, i.e., zero tangential electric field on the cylinder
boundary.
\end{proof}
}

\section{Theoretical Framework}
\label{sec:Theoretical Framework}
We show the derivation of the EM fields scattered from a single dielectric cylinder under excitation by: i) a line-source current; ii) an infinitesimal dipole. The two scenarios constitute core components of the ISAC channel model and form fundamental elements for developing the MIMO channel matrix in the presence of deterministic scatterers. Importantly, the line-source case can be captured in full by a 2D configuration and is well suited for objects whose size is comparable with the excitation extent. The dipole source illuminates only part of the infinitely extended cylinder and thus constitutes a crucial step towards 3D modelling. The model is suitable for objects whose dimensions are larger than the excitation extent.
\subsection{Line-source excitation}
\label{subsec:lineexcitation}
Figure 1(a) illustrates the radio wave propagation between the
transmitter and the receiver influenced by the
presence of the scatterer $\S$. Let $\T$ be an electric current filament with a time-harmonic variation of $e^{j \omega t}$ and a phasor of $\It$, located at $\rt = [\at,\bt] = \at \hat{\bf x} + \bt \hat{\bf y} $ where $\hat{\bf x}$ and $\hat{\bf y}$ are the unit vectors in the directions of $\emph{x}$ and $\emph{y}$ axes, respectively. Let $\R$ be a receiver located at $\rr_{\rm r} = [\ar , \br]$.  Let a scatterer, i.e., the target, $\S$ be present in the environment and located at $\rs = [\as , \bs]$. The scatterer is assumed to
be an infinite dielectric cylinder of radius $a$ whose relative permittivity in the frequency band of interest is known to be a complex function of frequency $\epsd(\omega)$. Let ${\rho}_{\mathsf{{e}_{1}}\mathsf{{e}_{2}}} =\|\boldsymbol{\rho}_{\mathsf{{e}_{1}}}-\boldsymbol{\rho}_{\mathsf{{e}_{2}}}\|$, where $\mathsf{{e}_{1}}$, $\mathsf{{e}_{2}} \in  \{ \r,\t,\s \}$ be the distance between two entities $\mathsf{{e}_{1}}$ and $\mathsf{{e}_{2}}$. Let also  $\boldsymbol{{\rho}}_{\mathsf{{e}_{1}}\mathsf{{e}_{2}}}  =\boldsymbol{\rho}_{\mathsf{{e}_{2}}}-\boldsymbol{\rho}_{\mathsf{{e}_{1}}}$ be the vector between the two entities. This scenario is shown in Fig.~\ref{fig:Config}.

Now, assume that the transmitter is an \emph{electric} current filament with a time-harmonic variation of $e^{j \omega t}$ and a phasor of $\It$ as can be seen in Fig.~\ref{fig:Config}.  One can readily show that in this two-dimensional case, the total electromagnetic field everywhere is in a $\text{TM}_z$ mode. The auxiliary scalar potential associated with the total
electric field at an arbitrary location measured from
the scatterer, $\S$, i.e., at a distance $\rho$ and angle
$\phi$ as in Fig.~\ref{fig:Config}, can be written as  
\begin{equation}
 \psi(\rho,\phi) =\psi_{\textrm{i}}(\rho,\phi) + \psi_{\textrm{s}}(\rho,\phi),
 \label{eq:totalphi}
\end{equation}
where the first term is due to the line of sight (LoS)
contribution of the transmitter, i.e., the incident
auxiliary scalar potential.\footnote{Note that when we
are interested in computing the field at the location
of the receiver, the generic angle $\phi$ can be
computed as ${\mathrm{arctan}}(\br-\bs,\ar-\as)$
and the distance $\rho$ will be $\rsr$.}
The corresponding electric field is linearly polarised
along the \emph{z}-axis and is related to the auxiliary
scalar potential by $E_z=-j\omega\muo\psi$.
The incident potential is calculated in the absence
of the scatterer $S$ and depends only on the spacing
between the receiver and the transmitter, which can
be written as~\cite{Harrington2001}
 \begin{align}
     \psi_{\textrm{i}}(\rho,\phi)
     &\triangleq\frac{\It}{4j}
     H_{0}^{(2)} (\ko |\boldsymbol{\rho}_{\text{ts}}+\boldsymbol{\rho}|),
     \label{eq.incidentlinesource}
\end{align}
where $\boldsymbol{\rho}$ is the vector between the scatterer and the arbitrary point, $|\boldsymbol{\rho}_{\text{ts}}+\boldsymbol{\rho}|$ is the distance between the transmitter and the arbitrary point in space,   $H_{0}^{(2)}$ represents the Hankel function of the second type and the zero-th order, $\omega$, $\muo$, $\epso$, $\ko\triangleq \omega \sqrt{\muo \epso}$ denote the angular operating frequency, permeability, permittivity, wave number, all in free space, respectively. To represent it in terms of $\rho$, we rewrite it as\cite{Harrington2001} 
\begin{align}
     \!\psi_{\textrm{i}}(\rho,\phi)
     &=\frac{\It}{4j}\sum_{n=-\infty}^{\infty} H_{n}^{(2)} (\ko \rts) J_{n} (\ko  \rho ) e^{jn(\phi - \phi')},
     \label{eq.incidentline}
 \end{align}
By the addition theorem for Hankel functions, we have
 \begin{align}
     \!\!H_{0}^{(2)} (\ko |\boldsymbol{\rho}_{\text{ts}}+\boldsymbol{\rho}|)
     &\!=\!\!\!\sum_{n=-\infty}^{\infty} \!\!H_{n}^{(2)} (\ko \rts) J_{n} (\ko  \rho ) e^{jn(\phi - \phi')}.
     \label{eq:graf_line_addition}
 \end{align}
where $J_n$ stands for the Bessel function of the first kind and the \emph{n}-th order, and $\phi'\triangleq \mathrm{atan2}(\bt-\bs,\at-\as)$ is the transmitter azimuth defined in Fig.~\ref{fig:Config}, i.e., $\phi'=\phi_t$. Here, $\mathrm{atan2}$ denotes the quadrant-aware inverse tangent function. 

The second term in \eqref{eq:totalphi}, $ \psi_{\textrm{s}}(\rho,\phi)$, is due to the scatterer and is calculated after solving the full-wave boundary-condition problem at
hand. To this end, the total $\text{TM}_z$ field inside and outside the scatterer is obtained by solving the Helmholtz wave equation in cylindrical coordinates whose origin is on the axis of the
dielectric cylinder, yielding\cite{Harrington2001} 
\begin{equation}
     \psi_{\textrm{s}}(\rho,\phi) =  \frac{1}{4j}
  \begin{cases} 
   \sum_{n=-\infty}^{\infty} I_n\, J_n (\kd \rho) e^{jn(\phi - \phi')} & \text{if } \rho \leq a \\
   \sum_{n=-\infty}^{\infty}\! I_{n}^{\textrm{s}} H_{n}^{(2)} (\ko \rho) e^{jn(\phi - \phi')}       & \text{if } \rho > a
  \end{cases}
  \label{eq.scatteredline}
\end{equation}
Moreover, the wave-number inside the dielectric cylinder is 
$\kd = \omega \sqrt{\muo \epso \epsd}$, and $I_n$ and $I^{s}_n$ are 
the unknown expansion coefficients to be determined in order to fulfil the boundary conditions. 

Now, imposing boundary conditions on the tangential electric fields, i.e., 
\begin{eqnarray}
    \psi_{\textrm{i}}(a^{+},\phi)+\psi_{\textrm{s}}(a^{+},\phi)=\psi_{\textrm{s}}(a^{-},\phi),
    \label{eq:contfields}
\end{eqnarray}
and the tangential magnetic fields 
yields the following system of equations for the unknown modal coefficients
\begin{multline}\label{eqn:FieldContFil}
  \begin{bmatrix}
   J_{n} (\kd \, a ) &
   - H_{n}^{(2)} (\ko a ) \\
   {\frac{\kd}{\ko} \, J'_{n} (\kd \, a )}  &
   - H_{n}^{(2)'} (\ko \, a ) 
   \end{bmatrix} 
   \begin{bmatrix} I_n \\ I_{n}^{\textrm{s}} \end{bmatrix}
    \\= \begin{bmatrix} J_{n} (\ko \, a ) \\ J'_{n} (\ko \, a ) 
   \end{bmatrix} \, H_{n}^{(2)} (\ko \rts ) \, \It, 
\end{multline}
which is valid for $n\in \{0,\pm 1,\pm 2, \cdots\}$, as detailed in Appendix~\ref{apendix:boundrycondlinesource}.\footnote{The magnetic fields are the derivatives of the electric fields with respect to $\rho$. Hence, the boundary conditions can be found following a procedure similar to \eqref{eq:contfields}.}
{
Eliminating the internal-field coefficient $I_n$ from \eqref{eqn:FieldContFil} gives
\begin{equation}
I_n^{\textrm{s}}=\Gamma_{d,n}(\omega,a,\epsd)
H_n^{(2)}(\ko\rts)\It,
\label{eq:In_general}
\end{equation}
where the complete modal scattering coefficient is
\begin{equation}
\Gamma_{d,n}=\frac{\dfrac{\kd}{\ko}J_n'(\kd a)J_n(\ko a)-J_n(\kd a)J_n'(\ko a)}{J_n(\kd a)H_n^{(2)'}(\ko a)-\dfrac{\kd}{\ko}J_n'(\kd a)H_n^{(2)}(\ko a)}.
\label{eq:Gamma_dn_general}
\end{equation}
Substituting \eqref{eq:In_general} into \eqref{eq.scatteredline} and adding the incident field yields the infinite modal expansion of the line-source field; following the standard modal-truncation practice for cylindrical scattering expansions~\cite{balanis2012advanced}, the numerical form summarized in \eqref{eqn:Ez_line} uses the symmetric truncated sum $\sum_{n=-N}^{N}$, with $N$ chosen to ensure convergence of the modal sum.}

{
When $a\ll\lambda_{\mathsf d}/(2\pi)$, or equivalently $|\kd|a\ll1$, the zeroth-order scattered coefficient is dominant~\cite{balanis2012advanced}, i.e.,
\begin{equation}
\left|I_0^{\textrm{s}}\right|\gg
\left|I_n^{\textrm{s}}\right|,\qquad n\neq0.
\end{equation}
Setting $n=0$ in \eqref{eq:Gamma_dn_general} and using $J_0'(x)=-J_1(x)$ and $H_0^{(2)'}(x)=-H_1^{(2)}(x)$ gives
the zeroth-order coefficient as}
\begin{equation}
\begin{split}
    I_{0}^{\textrm{s}} = \frac{-J_{0} (\kd \, a ) J_{1} (\ko \, a ) + \sqrt{\epsd} J_{0} (\ko \, a ) J_{1} (\kd \, a )}{J_{0} (\kd \, a ) H_{1}^{(2)} (\ko a ) - \sqrt{\epsd} J_{1} (\kd \, a ) H_{0}^{(2)} (\ko \, a )} \\ \times H_{0}^{(2)} (\ko \rts ) \, \It. 
\end{split}
\end{equation}
This last relation can be expressed as
\begin{equation}
    I_{0}^{\textrm{s}} = \Gamma_d (\omega, a, \epsd)\, H_{0}^{(2)} (\ko \rts ) \, \It, 
    \label{eq:I0ssmallr}
\end{equation}
{where the zeroth-order complex modal scattering coefficient is}
\begin{equation}
    \Gamma_d (\omega, a, \epsd) = \frac{-J_{0} (\kd \, a ) J_{1} (\ko \, a ) + \sqrt{\epsd} J_{0} (\ko \, a ) J_{1} (\kd \, a )}{J_{0} (\kd \, a ) H_{1}^{(2)} (\ko a ) - \sqrt{\epsd} J_{1} (\kd \, a ) H_{0}^{(2)} (\ko \, a )} 
\end{equation}
{and depends on the material properties and size of the dielectric cylinder, as well as the operating frequency.}

Starting from $I_{0}^{\textrm{s}}$, one can determine the total electric field 
at the location of the receiver  as the sum of two terms, i.e.,
\begin{equation}
    E_z \left ( \ar, \br \right ) \approx -\frac{\omega \muo}{4} 
    \left ( \It H_{0}^{(2)} (\ko \, \rtr ) + I_{0}^{\textrm{s}} H_{0}^{(2)} (\ko \, \rsr ) \right ), 
\end{equation}
in which the first term represents the exact LoS contribution whereas
the second term represents an accurate approximation of the electric
field scattered by the dielectric cylinder and $I_{0}^{\textrm{s}}$ is given by \eqref{eq:I0ssmallr}. As we will demonstrate, this
is highly accurate for cylinders of small radius with respect to the
wavelength inside the cylinder.

\subsection{Dipole excitation}
\label{subsec:dipole}

We now consider a localised source of excitation that is localised in a (three-dimensional) space at $\rt$ and is expressed by the following dipole current
\begin{equation}
 \textbf{J}_d = I^t \, dl \, \delta_{3} \left ( \ro - \rt\right ) \, \hat{\bf z} , 
\end{equation}
where $\ro \triangleq [x,y,z]$ is the 3D vector from the origin to the point in space indicated by the coordinates $[x,y,z]$,  $I^t$ is the current phasor, $dl$ is an infinitesimal dipole length along the $\hat{\bf z}$ direction, and  $\delta_{N} (\cdot)$ represents a Dirac delta in $N$ dimensions.
Suppose the dipole excites a dielectric cylinder as shown in Fig.~\ref{fig:Config}.
Unfolding the Dirac delta in Cartesian components and using the integral representation of the Dirac delta along the $\hat{\bf z}$  direction yields
\begin{equation}
    \textbf{J}_d = I^t \, dl \, \delta \left ( x - x_0\right ) \, \delta \left ( y - y_0\right )  \, \frac{1}{2 \pi} \int_{-\infty}^{\infty} e^{-j k_z z} d k_z \, \hat{\bf z} ,
\end{equation}
where $k_z$ is the wave-number along $\hat{\bf z}$. 
Considering the infinitesimal current filament 
\begin{equation}
    d\textbf{J}_d = I^t \, dl \, \delta \left ( x - x_0\right ) \, \delta \left ( y - y_0\right )  \, \frac{d k_z}{2 \pi} \, e^{-j k_z z} \, \hat{\bf z}.
    \label{eq.dj}
\end{equation}

 In order to derive the incident field radiated by the current density, 
we solve the Helmholtz equation in cylindrical coordinates\cite[Sec.~9.2]{balanis2012advanced}. For a fixed spectral component $k_z$, the corresponding $\mathrm{TM}_z$ scalar field 
$\psi_i(\rho,\phi;k_z)$ satisfies a two-dimensional Helmholtz equation in the transverse plane,
\begin{equation}
\left(\nabla_t^2 + k_\rho^2\right)\psi_i(\rho,\phi;k_z)
=
-\delta(\boldsymbol{\rho}-\boldsymbol{\rho}_t),
\end{equation}
where $\boldsymbol{\rho}=(\rho,\phi)$ denotes the observation point and 
$\boldsymbol{\rho}_t=(\rho_t,\phi_t)$ represents the transverse location of the dipole source, 
with $\rho_t=\sqrt{x_0^2+y_0^2}$ and $\phi_t=\operatorname{atan2}(y_0,x_0)$.
Here $k_\rho$ is the transverse wavenumber defined by $
k_\rho^2 \triangleq k_0^2 - k_z^2 $,
with $k_0$ being the free-space wavenumber.

The outgoing-wave Green’s function solution of the above equation is given by
\begin{equation}
G(\boldsymbol{\rho},\boldsymbol{\rho}_t)
=
\frac{1}{j4}\,
H_0^{(2)}\!\left(
k_\rho \left|\boldsymbol{\rho}-\boldsymbol{\rho}_t\right|
\right).
\end{equation}

For a fixed spectral component $k_z$, the infinitesimal current filament
\eqref{eq.dj} gives rise to a transverse point source
$\delta(\boldsymbol{\rho}-\boldsymbol{\rho}_t)$ in the $x$–$y$ plane.
Accordingly, the associated $\mathrm{TM}_z$ scalar field
$\psi_i(\boldsymbol{\rho};k_z)$ admits the Green’s function representation
\begin{equation}
\psi_i(\boldsymbol{\rho};k_z)
=
\iint_{\mathbb{R}^2}
G(\boldsymbol{\rho},\boldsymbol{\rho}')
\delta(\boldsymbol{\rho}'-\boldsymbol{\rho}_t)\,
d^2\boldsymbol{\rho}',
\end{equation}
which, upon evaluation of the delta function, yields
$\psi_i(\boldsymbol{\rho};k_z)=G(\boldsymbol{\rho},\boldsymbol{\rho}_t)$.
The corresponding three-dimensional field contribution is then obtained
via inverse Fourier transformation along $z$.

Accordingly, the incident field contribution associated with the current filament \eqref{eq.dj} can be written as
\begin{equation}
d\psi_i(\rho,\phi,z)
=
\frac{1}{j4}
\left(
I^t dl \,\frac{dk_z}{2\pi}
\right)
e^{-j k_z z}
H_0^{(2)}\!\left(
k_\rho \left|\boldsymbol{\rho}-\boldsymbol{\rho}_t\right|
\right).
\label{eq:dpsi_h0}
\end{equation}

To express \eqref{eq:dpsi_h0} in a form centered at the cylinder axis, 
Graf’s addition theorem is applied. For $\rho \le \rho_t$, the Hankel function admits the cylindrical-wave expansion
\begin{equation}
\!\!H_0^{(2)}\!\left(
k_\rho \left|\boldsymbol{\rho}-\boldsymbol{\rho}_t\right|
\right)
\!=\!
\!\!\!\sum_{n=-\infty}^{\infty}
\!H_n^{(2)}(k_\rho \rho_t)\,
J_n(k_\rho \rho)\,
e^{j n (\phi-\phi_t)} .
\end{equation}
Substituting the above expansion into \eqref{eq:dpsi_h0}, 
        the incident $\mathrm{TM}_z$ field at $(\rho,\phi,z)$ from the center of the cylinder is obtained as
\begin{multline}
\label{eq:fieldincidentdipole}
     d\psi_{\textrm{i}} \left ( \rho, \phi, z \right ) = \frac{1}{j 4} \left ( I^t \, dl \, \frac{d k_z}{2 \pi} \right ) \, e^{-j k_z z} \\ \times\sum^{\infty}_{n=-\infty} H_{n}^{(2)} (k_{\rho} \, \rts ) \, J_{n} (k_{\rho} \, \rho ) \, e^{j n ( \phi - \phi')} , \,\,\,\,\,\,\, \rho \leq \rts ,
\end{multline}
where $\phi'$ is the {AoA} at the cylinder. 
The related scattered $\text{TE}_z$ field yields 
\begin{multline}
\label{eq:dfieldsoutside}
 d\psi_{\textrm{s}}^{\textrm{e}}\left ( \rho, \phi, z \right ) =  \frac{1}{j 4} \left ( dl \, \frac{d k_z}{2 \pi} \right ) \, e^{-j k_z z} \\ \times \sum^{\infty}_{n=-\infty} K_n^s \, H_{n}^{(2)} (k_{\rho} \, \rho ) \, e^{j n ( \phi - \phi')} ,  \,\,\,\,\,\,\, \rho \geq a ,
\end{multline}
\begin{multline}
\label{eq:dfieldinsidedipole}
     d\psi_{\textrm{1}}^{\textrm{e}} \left ( \rho, \phi, z \right ) =  \frac{1}{j 4} \left ( dl \, \frac{d k_z}{2 \pi} \right ) \, e^{-j k_z z} \\ \times\sum^{\infty}_{n=-\infty} K_n \, J_{n} (k_{\mathsf{\rho d}} \, \rho ) \, e^{j n ( \phi - \phi')} , 
    \,\,\,\,\,\,\, \rho \leq a ,
\end{multline}
where $k^2_{\mathsf{\rho d}}+k_z^2=\kd^2=\ko^2\epsd$. The related scattered $\text{TM}_z$ field yields
\begin{equation} \label{eqn:dpsims}
\begin{split}
     d\psi_{\textrm{s}}^{\textrm{m}} \left ( \rho, \phi, z \right ) =  \frac{1}{j 4} \left ( dl \, \frac{d k_z}{2 \pi} \right ) \, e^{-j k_z z} \\ \sum^{\infty}_{n=-\infty} I_{n}^{\textrm{s}} \, H_{n}^{(2)} (k_{\rho} \, \rho ) \, e^{j n ( \phi - \phi')} ,  \,\,\,\,\,\,\, \rho \geq a ,
\end{split}
\end{equation}
\begin{equation}
\label{eqn:dpsimsi}
\begin{split}
   d\psi_{\textrm{1}}^{\textrm{m}} \left ( \rho, \phi, z \right ) =  \frac{1}{j 4} \left ( dl \, \frac{d k_z}{2 \pi} \right ) \, e^{-j k_z z} \\ \sum^{\infty}_{n=-\infty} I_n \, J_{n} (k_{\mathsf{\rho d}} \, \rho ) \, e^{j n ( \phi - \phi')} , 
    \,\,\,\,\,\,\, \rho \leq a ,
\end{split}
\end{equation}
Now, imposing the continuity of the tangential electric fields at $\rho=a$ across the cylinder boundary as described in Appendix~\ref{apendix:boundrycondlinesource} and rearranging it in a matrix form gives four independent equations for the four unknown $\text{TE}_z$ and $\text{TM}_z$ modal coefficients, i.e.,  $K_n^s$, $K_n$, $I_{n}^{\textrm{s}}$, $I_n$. This system of equations is represented as in \eqref{eqn:FieldCont}.  
\begin{figure*}
\begin{gather}\label{eqn:FieldCont}
 \!\!\!\!\!\! \underbrace{\begin{bmatrix}
   -H_{n}^{(2)} (k_{\rho} \, a ) &
   0 & 
   \frac{k^2_{\rho d}}{\epsd k^2_{\rho}} \, J_{n} (k_{\mathsf{\rho d}} \, a ) &
   0 \\
   j \frac{k_z}{\omega \epso} \frac{n}{a} \, H_{n}^{(2)} (k_{\rho} \, a )  &
   - k_{\rho} \, H_{n}^{(2)'} (k_{\rho} \, a ) & 
   - j \frac{k_z}{\omega \epso \epsd} \frac{n}{a} \, J_{n} (k_{\mathsf{\rho d}} \, a ) & 
   k_{\mathsf{\rho d}} \, J_{n}^{'} (k_{\mathsf{\rho d}} \, a ) \\
   0 &
   - H_{n}^{(2)} (k_{\rho} \, a ) &
   0 &
   \frac{k^2_{\rho d}}{k^2_{\rho}} \, J_{n} (k_{\mathsf{\rho d}} \, a ) \\
   -H_{n}^{(2)'} (k_{\rho} \, a ) &
   -j \frac{k_z}{\omega \muo} \frac{n}{k_{\rho} a} \, H_{n}^{(2)} (k_{\rho} \, a ) & 
   \frac{k_{\mathsf{\rho d}}}{k_{\rho}} \, J'_{n} (k_{\mathsf{\rho d}} \, a ) &
   j \frac{k_z}{\omega \muo} \frac{n}{k_{\rho} a} \, J_{n} (k_{\mathsf{\rho d}} \, a )
   \end{bmatrix}}_{\mathbf{A}} \!\!\!
   \begin{bmatrix} I_{n}^{\textrm{s}} \\ K_n^s \\ I_n \\ K_n \end{bmatrix}
\!\!=\!\! \underbrace{\begin{bmatrix} J_{n} (k_{\rho} \, a ) \\ -j \frac{k_z}{\omega \epso} \frac{n}{a} \, J_{n} (k_{\rho} \, a ) \\ 0 \\ J'_{n} (k_{\rho} \, a ) \end{bmatrix}}_{\mathbf{c}}  \!\!H_{n}^{(2)} (k_{\rho} \, \rts )  I^t, 
\end{gather}
\end{figure*}
We have the following recurrence rules
\begin{equation}\label{eqn:Jrecurr}
 J'_{n} (k_{\mathsf{\rho d}} \, a ) = - J_{n+1} (k_{\mathsf{\rho d}} \, a ) + \frac{n}{k_{\mathsf{\rho d}} \, a} J_{n} (k_{\mathsf{\rho d}} \, a ),
\end{equation}
\begin{equation}\label{eqn:H2recurr}
 H^{(2)'}_{n} (k_{\rho} \, a ) = - H^{(2)}_{n+1} (k_{\rho} \, a ) + \frac{n}{k_{\rho} \, a} H^{(2)}_{n} (k_{\rho} \, a ),
\end{equation}
for every $\ko, k_z, n, a, \epsd$,
where $ J'_{n}(x)$ and $H^{(2)'}_{n}(x)$ are the derivatives of $ J_{n}(x)$ and $H^{(2)}_{n}(x)$ with respect to $x$.
Note that $I^{s}_n \equiv I^{s}_n (k_z)$ for given $\omega, a, \epsd$, which is valid for any modal coefficient type (i.e. induced currents) with the symmetry properties described in Appendix \ref{appendix:bondatconddipole}. 
We are now ready to determine the total scattered field at $\rho > a$ (i.e., outside the target body), perceived by a receiving dipole polarized along $\hat{\mathbf{z}}$. 
This is done by integrating the infinitesimal electric field,
\begin{equation}\label{eqn:dEz}
 dE_z = dE^{i}_z + dE^{s}_z = dE^{i}_z + \frac{1}{j \omega \epso} \, \left ( \frac{\partial^2}{\partial z^2} + \ko^2 \right ) \, d\psi_s^m,
\end{equation}
over $k_{z}$. By substituting (\ref{eqn:dpsims}) in (\ref{eqn:dEz}) and noting that the receiver is at distances $\rsr$ and $\rtr$ from the target and the transmitter, respectively, we get the total electric field
\begin{align}
E_z(\rho,\phi,z)
&= \int dE_z
= E_z^{\rm i}(\rho_{tr})
-\frac{dl}{8\pi\omega\epsilon_0}
\int_{-\infty}^{\infty}
k_\rho^2(k_z)e^{-jk_z z}\nonumber\\
&\quad\times
\sum_{n=-\infty}^{\infty}
I_n^{\rm s}(k_z)
H_n^{(2)}\left(k_\rho(k_z)\rho\right)
e^{jn(\phi-\phi')}\,dk_z.
\label{eq:dipole_total_field}
\end{align}
where the incident field is readily found by the field
radiated by a dipole in the plane $\theta=\frac{\pi}{2}$,
corresponding to $z=0$, as
\begin{equation}\label{eqn:Ezi_final}
 E^{i}_z(\rtr) = - j \omega \muo I^t dl \frac{e^{-j \ko \rtr}}{4 \pi \rtr} \left ( 1 - j \frac{1}{\ko \rtr} -\frac{1}{\left ( \ko \rtr \right )^2} \right ),
\end{equation}
which is detailed in Appendix~\ref{appendix:bondatconddipole}.
The induced scattering field can be computed from \eqref{eqn:FieldCont} as
\begin{align}
  I_{n}^{\textrm{s}}(k_z)&=  {[{\mathbf{A}^{-1}_{n}(k_z)}\, \mathbf{c}_{n}(k_z)]}_{1}\, I^{t} H_{n}^{(2)} (k_{\rho} \, \rts ) \nonumber \\ 
  &\triangleq \Gamma_n (k_z)\, I^{t} H_{n}^{(2)} (k_{\rho} \, \rts ),
\end{align}
{where $\Gamma_n(k_z)$ is the complex modal scattering coefficient associated with $I_n^{\textrm{s}}(k_z)$.}
{Using the modal scattering coefficient, the total electric field evaluated on the $z=0$ plane is approximated by the following truncated coherent modal sum}
{
\begin{multline}
\label{eqn:Eztot}
    E_z (\rsr) \simeq E^{i}_z(\rtr) - \frac{\omega \muo I^{t} dl}{8 \pi} \,
    \\ \times 
    \sum^{N}_{n=-N} e^{j n ( \phi - \phi')}
    \int^{\infty}_{-\infty} \!\left(1- \frac{k_{z}^2}{\ko^2} \right ) \, \Gamma_n (k_z)\,
    \\ \times
    H_{n}^{(2)} \Big(k_{\rho}(k_z) \, \rts \Big) \, H_{n}^{(2)} \Big(k_{\rho}(k_z) \, \rsr \Big) \, dk_z.
\end{multline}
}
{For an electrically small cylinder, $|\kd|a\ll1$, the zeroth-order mode gives}
\begin{equation}
\begin{split}
    E_z(\rsr) \approx E^{i}_z(\rtr) - \frac{\omega \muo I^{t} dl}{8 \pi} \, \int^{\infty}_{-\infty} \,  \left(1- \frac{k_{z}^2}{\ko^2} \right ) \, \\ \Gamma_0 (k_z) \, H_{0}^{(2)} (k_{\rho} \, \rts ) \, H_{0}^{(2)} (k_{\rho} \, \rsr ) \, dk_z.
\end{split}
\end{equation}
\medskip

The channel transfer function, $h_{\mathsf{tr}}$, can be easily found from the electric field. For instance, if the receiver adopts a small dipole, then the trans-impedance (equivalent to the channel model) can be written as
\begin{equation}
   h_{\mathsf{tr}}\triangleq Z_{\mathsf{tr}} = \frac{V_{\mathsf{r}}}{I^t}, 
\end{equation}
where $V_{\mathsf{r}}= -E_z dl$ is the open-circuit voltage for the receiving dipole.

\section{Sensing Performance with the Proposed EM Model}
In this section, we formulate the SIMO sensing observation model and derive the \ac{CRLB} for target-position estimation.
\label{Sec:Sensing}
\subsection{Signal Model with line excitation}
\label{subsec:CRLB}

Although the electromagnetic field expression has been derived in Section~\ref{sec:Theoretical Framework}, we now recast it as a SIMO sensing observation model in order to formulate the likelihood function and the \ac{FIM}. Let the unknown target-position vector be
\begin{equation}
\mathbf{s}\triangleq[x,y]^{\mathsf T}.
\label{eq:position_vector}
\end{equation}

{
For a fixed transmitter or the $b$-th receiving antenna, indexed by $e\in\{t,b\}$, define
\begin{equation}
\begin{aligned}
\mathbf d_e&\triangleq\mathbf r_e-\mathbf s
=[d_{e,x},d_{e,y}]^{\mathsf T},\\
\rho_e&\triangleq\|\mathbf d_e\|,
\qquad
\phi_e\triangleq\operatorname{atan2}(d_{e,y},d_{e,x}).
\end{aligned}
\label{eq:full_modal_geometry}
\end{equation}
For $e=t$, $\rho_e$ and $\phi_e$ represent the range and azimuth from the target to the transmitter; for $e=b$, they represent the range and azimuth from the target to the $b$-th receiving antenna.
We consider an array of $\Nb$ receiving antennas with centre at $[\ar,\br]$.}

The induced noisy electric field at the location of the $b$-th receiving antenna can be rewritten as 
\begin{align}\label{eq:recsigsum}
\yb=& \mu_{b}(\mathbf{s})  + \nb  \,, &&\forall\,  b \in\{ 1,2, \cdots, \Nb\},
\end{align}
where $\mu_b(\mathbf{s})$ is the noise-free received electric field at the $b$-th receiving antenna obtained from \eqref{eqn:Ez_line}. Defining $\mathbf y=[y_1,\ldots,y_{\Nb}]^{\mathsf T}$ and $\mathbf n=[n_1,\ldots,n_{\Nb}]^{\mathsf T}$, we assume independent proper complex Gaussian noise,
\begin{equation}
\mathbf n\sim\mathcal{CN}\!\left(\mathbf 0,
\boldsymbol{\Sigma}\right),\qquad
\boldsymbol{\Sigma}=\operatorname{diag}(\sigma_1^2,\ldots,\sigma_{\Nb}^2),
\end{equation}
where $\boldsymbol{\Sigma}$ does not depend on the unknown target position $\mathbf{s}$.

{
To emphasize the dependence of the complete modal field on the target position, \eqref{eqn:Ez_line} is written for the $b$-th receiver as
\begin{align}
\mu_b(\mathbf{s})
={}&\mu_b^{\mathrm{LoS}}
+\alpha\sum_{n=-N}^{N}g_{b,n}(\mathbf{s}),
\label{eq:full_modal_mean}\\
\mu_b^{\mathrm{LoS}}
\triangleq{}&\alpha H_0^{(2)}(\ko\rtrb),
\qquad
\alpha\triangleq-\frac{\omega\muo\It}{4},
\end{align}
where $\rtrb$ is the transmitter--receiver distance and
\begin{equation}
g_{b,n}(\mathbf{s})
\triangleq
\Gamma_{d,n}
H_n^{(2)}(\ko\rts)
H_n^{(2)}(\ko\rb)
e^{jn(\phi_b-\phi_t)}.
\label{eq:full_modal_kernel}
\end{equation}
The cylinder radius, material properties, and operating frequency are assumed known, so $\Gamma_{d,n}$ is independent of the unknown target position $\mathbf{s}$. The direct transmitter--receiver term $\mu_b^{\mathrm{LoS}}$ is also independent of $\mathbf{s}$.}
\subsection{Localization Algorithm}
~
One possible solution for estimating the target position is the maximum-likelihood estimator (MLE). Using the position vector defined in \eqref{eq:position_vector}, the MLE maximizes the likelihood of the complete receive-array observation~\cite{ElzanatyAnnaFrans:21}, i.e.,
\begin{equation}
\widehat{\mathbf{s}} \triangleq \underset{\mathbf{s} \in \mathbb{R}^{2}}{\mathrm{argmax}} \log p(\mathbf y; \mathbf{s}).
\end{equation}

This optimization problem can be solved using grid search or iterative algorithms such as Newton-Raphson, though it should be noted that iterative methods may not guarantee convergence to the global maximum.

Of particular theoretical importance is the asymptotic behavior of the MLE and its connection to the FIM. Under standard regularity conditions and in the asymptotic regime of high signal-to-noise ratio or a sufficiently large number of observations, the MLE is asymptotically efficient, and its estimation error is approximately distributed as
\begin{equation}
\widehat{\mathbf{s}}-\mathbf{s}
\overset{\mathrm{a}}{\sim}
\mathcal{N}\!\left(\mathbf{0},\boldsymbol{\Lambda}(\mathbf{s})\right),
\end{equation}
{where $\boldsymbol{\Lambda}(\mathbf{s})$ denotes the inverse of the total FIM}, i.e., the CRLB matrix. Hence, the error covariance of the MLE asymptotically approaches the CRLB. Therefore, in the following, we derive the CRLB to characterize the fundamental performance limit of unbiased target-position estimators. The corresponding matched-model MLE and position error bound are compared numerically in Section~\ref{Sec:NumericalResults}.

\subsection{CRLB for position}

\begin{figure}[t!]
    \centering
\includegraphics[width=0.6\linewidth]{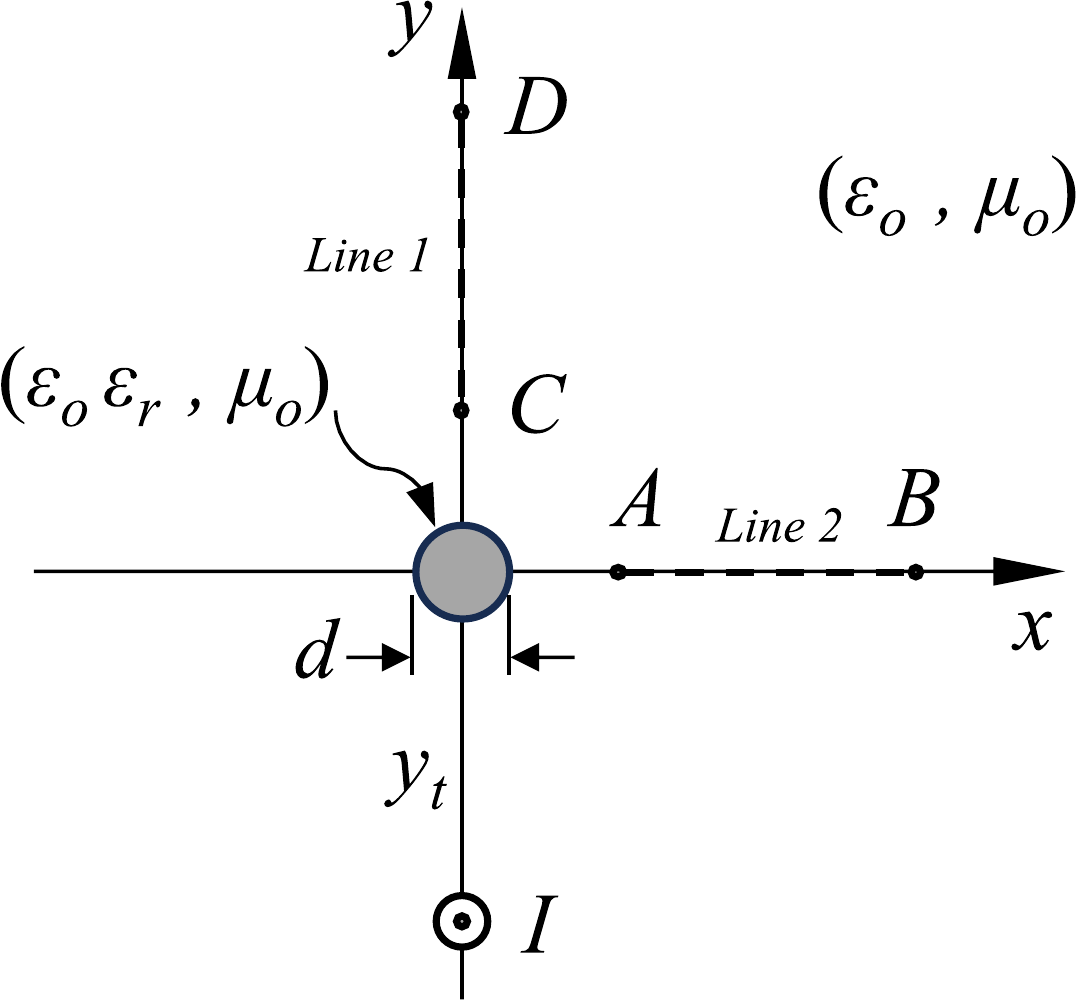}
    \caption{The simulation scenario considered for both COMSOL simulations and analytical results in Fig.~\ref{fig:CombinedELine1} and Fig.~\ref{fig:CombinedELine2}, depicting a line excitation at $(0, \bt)$, a target at the origin, and the electric field computed along 'Line 1' (i.e., $\ar=0$, and varying $\br$ between points C and D) or 'Line 2' (i.e., $\br=0$, while varying $\ar$ between points A and B).}
    \label{fig:ISAC_er_COMSOL}
\end{figure}

{
We now differentiate the complete modal mean in \eqref{eq:full_modal_mean} with respect to the position vector $\mathbf{s}$ defined in \eqref{eq:position_vector} and use the resulting derivatives to construct the FIM. Since the field depends on the target position through the ranges and azimuth angles in \eqref{eq:full_modal_geometry}, the required Cartesian geometry derivatives are
\begin{align}
\frac{\partial\rho_e}{\partial x}
&=-\frac{d_{e,x}}{\rho_e},
&
\frac{\partial\rho_e}{\partial y}
&=-\frac{d_{e,y}}{\rho_e},
\label{eq:full_modal_range_derivatives}\\
\frac{\partial\phi_e}{\partial x}
&=\frac{d_{e,y}}{\rho_e^2},
&
\frac{\partial\phi_e}{\partial y}
&=-\frac{d_{e,x}}{\rho_e^2}.
\label{eq:full_modal_angle_derivatives}
\end{align}}

\begin{figure}[t!]
    \centering
    \begin{subfigure}[b]{0.48\textwidth}
        \centering
        \pgfplotsset{every axis/.append style={
		font=\footnotesize,
		line width=1pt,
		legend style={font=\footnotesize, at={(1,1)}},legend cell align=left},
} \pgfplotsset{compat=1.13}
	\begin{tikzpicture}\begin{axis}[
xlabel near ticks,
grid=major,
xlabel={$\br$ (m)},
ylabel={$|E_{z}|$ (dBV/m)},
width=0.85\linewidth,yticklabel style={/pgf/number format},
legend entries={Analytical\text{,} $\bt=-3~\text{cm} = \lambda/2$, 
Large Arg. Approx.\text{,} $\bt=-3~\text{cm} =\lambda/2$,
	COMSOL Sim.\text{,} $\bt=-3~\text{cm} =\lambda/2$, Analytical\text{,} $\bt=-30$~cm,  
    COMSOL Sim.\text{,} $\bt=-30$~cm,
	},
	ymin=29, ymax=45, xmax=0.45,
ylabel style={font=\footnotesize},
xlabel style={font=\footnotesize},
]
\addplot[red,solid,mark=diamond,mark repeat=30,mark size=3] table {Figures/EMReLine1Absd3axisY.dat};
\addplot[
    yellow,dashed
] coordinates {
(0.02, 35.9979)
(0.0316705, 35.4492)
(0.043341, 35.0268)
(0.0550114, 34.6785)
(0.0666819, 34.3807)
(0.0783524, 34.1201)
(0.0900229, 33.888)
(0.101693, 33.6788)
(0.113364, 33.4882)
(0.125034, 33.3132)
(0.136705, 33.1514)
(0.148375, 33.0009)
(0.160046, 32.8602)
(0.171716, 32.7281)
(0.183387, 32.6036)
(0.195057, 32.486)
(0.206728, 32.3743)
(0.218398, 32.2682)
(0.230069, 32.167)
(0.241739, 32.0704)
(0.25341, 31.9778)
(0.26508, 31.8891)
(0.276751, 31.8038)
(0.288421, 31.7218)
(0.300092, 31.6428)
(0.311762, 31.5665)
(0.323432, 31.4929)
(0.335103, 31.4216)
(0.346773, 31.3527)
(0.358444, 31.2858)
(0.370114, 31.221)
(0.381785, 31.158)
(0.393455, 31.0968)
(0.405126, 31.0373)
(0.416796, 30.9794)
(0.428467, 30.923)
(0.440137, 30.868)
};
\addplot[gray!50!black,mark=x,mark repeat=30,mark size=3,mark options=solid, only marks] table {Figures/EMReLine1AbsSimd3axisY.dat};
\addplot[blue,solid,mark=diamond,mark repeat=30,mark size=3] table {Figures/EMReLine1Absd30axisY.dat};
\addplot[orange!20!brown,mark=x,mark repeat=30,mark size=3,mark options=solid, only marks] table {Figures/EMReLine1AbsSimd30axisY.dat};

\end{axis}
\end{tikzpicture}
        \vspace{-0.1cm}
        \caption{Absolute value of $E_\text{tot}$}
       \vspace{0.1cm}
        \label{fig:AbsELine1}
    \end{subfigure}
    \hfill
    \begin{subfigure}[b]{0.48\textwidth}
        \centering
        \pgfplotsset{every axis/.append style={
		font=\footnotesize,
		line width=1pt,
		legend style={font=\footnotesize, at={(0.99,0.99)}},legend cell align=left},
} \pgfplotsset{compat=1.13}
	\begin{tikzpicture}\begin{axis}[
xlabel near ticks,
ylabel near ticks,
grid=major,
xlabel={$\br$ (m)},
ylabel={$\operatorname{Re}\{E_z\}$ (V/m)},
width=0.85\linewidth,yticklabel style={/pgf/number format},
legend entries={Analytical\text{,} $\bt=-3~\text{cm} = \lambda/2$, COMSOL Sim.\text{,} $\bt=-3~\text{cm} =\lambda/2$, Analytical\text{,} $\bt=-30$~cm,  
    COMSOL Sim.\text{,} $\bt=-30$~cm,
	},
	ymin=-4000, ymax=5500,
ylabel style={font=\footnotesize},
xlabel style={font=\footnotesize},
]
\addplot[red,solid,mark=diamond,mark repeat=15,mark size=3] table {Figures/EMReLine1d3axisY.dat};
\addplot[gray!50!black,mark=x,mark repeat=15,mark size=3,mark options=solid, only marks] table {Figures/EMReLine1Simd3axisY.dat};
\addplot[blue,solid,mark=diamond,mark repeat=15,mark size=3] table {Figures/EMReLine1d30axisY.dat};
\addplot[orange!20!brown,mark=x,mark repeat=15,mark size=3,mark options=solid, only marks] table {Figures/EMReLine1Simd30axisY.dat};

\end{axis}
\end{tikzpicture}
        \vspace{-0.3cm}
        \caption{Real part of $E_\text{tot}$}
        \label{fig:RealELine1}
    \end{subfigure}
    \caption{Total received electric field along the y-axis (Line 1) with the target at the origin and $\ar=0$, for $\bt \in \{3, 30\}$~cm: (a) absolute value and (b) real part.}
    \label{fig:CombinedELine1}
\end{figure}
{
Since $\Gamma_{d,n}$ is position independent, the derivative of the $n$-th modal kernel in \eqref{eq:full_modal_kernel}, for $s_i\in\{x,y\}$, is
\begin{align}
\frac{\partial g_{b,n}}{\partial s_i}
={}&\Gamma_{d,n}e^{jn(\phi_b-\phi_t)}
\Bigg[
\ko H_n^{(2)'}(\ko\rts)
H_n^{(2)}(\ko\rb)
\frac{\partial\rts}{\partial s_i}
\nonumber\\
&+\ko H_n^{(2)}(\ko\rts)
H_n^{(2)'}(\ko\rb)
\frac{\partial\rb}{\partial s_i}
\nonumber\\
&+jnH_n^{(2)}(\ko\rts)
H_n^{(2)}(\ko\rb)
\left(
\frac{\partial\phi_b}{\partial s_i}
-\frac{\partial\phi_t}{\partial s_i}
\right)
\Bigg].
\label{eq:full_modal_kernel_derivative}
\end{align}
The Hankel derivative can be evaluated as
\begin{equation}
H_n^{(2)'}(z)
=\frac{1}{2}\left[
H_{n-1}^{(2)}(z)-H_{n+1}^{(2)}(z)
\right].
\label{eq:hankel_derivative_identity}
\end{equation}
Because the LoS term is independent of the target position, its derivative is zero. Hence,
\begin{equation}
\frac{\partial\mu_b}{\partial s_i}
=\alpha\sum_{n=-N}^{N}
\frac{\partial g_{b,n}}{\partial s_i},
\qquad s_i\in\{x,y\}.
\label{eq:full_modal_mean_derivative}
\end{equation}

For the independent proper complex Gaussian observations in \eqref{eq:recsigsum}, let $s_1=x$ and $s_2=y$. The FIM for the real position vector is given by~\cite[Ch.~2]{VanTrees:02}
\begin{equation}
\begin{aligned}
[\mathbf J(\mathbf s)]_{i,j}
=\sum_{b=1}^{\Nb}\frac{2}{\sigma_b^2}
\operatorname{Re}\!\left\{
\left(\frac{\partial\mu_b}{\partial s_i}\right)^*
\frac{\partial\mu_b}{\partial s_j}
\right\}.
\end{aligned}
\label{eq:full_modal_fim}
\end{equation}
Because \eqref{eq:full_modal_mean_derivative} is a coherent modal sum, \eqref{eq:full_modal_fim} automatically retains the cross terms between different modal orders.

The CRLB matrix and the PEB are finally
\begin{equation}
\boldsymbol{\Lambda}(\mathbf s)
=\mathbf J^{-1}(\mathbf s),
\qquad
\PEB
=\sqrt{\operatorname{tr}\!\left(
\boldsymbol{\Lambda}(\mathbf s)
\right)}.
\label{eq:full_modal_crlb_peb}
\end{equation}
For the electrically small-cylinder corollary, retaining only $n=0$ removes the angular term in \eqref{eq:full_modal_kernel_derivative}; using $H_0^{(2)'}(z)=-H_1^{(2)}(z)$ then recovers the zeroth-order derivatives.}

\subsection{Signal Model with dipole excitation}
{
For dipole excitation, let $\mu_b^{\mathrm{dip}}(\mathbf{s})$ denote the noiseless field in \eqref{eq:EVDipole} evaluated at the $b$-th receiving antenna. The direct dipole field depends only on the fixed transmitter--receiver geometry and therefore has zero derivative with respect to $\mathbf{s}$. For a converged modal sum and spectral integral, differentiation with respect to the transverse target coordinates can be interchanged with the spectral integration. Since $\Gamma_n(k_z)$ is independent of $\mathbf{s}$, the derivative of the scattered contribution is
\begin{multline}
\frac{\partial\mu_b^{\mathrm{dip}}}{\partial s_i}
=-\frac{\omega\muo I^t dl}{8\pi}
\sum_{n=-N}^{N}\int_{-\infty}^{\infty}
\left(1-\frac{k_z^2}{\ko^2}\right)
\Gamma_n(k_z)\\
\times\frac{\partial}{\partial s_i}
\left[
H_n^{(2)}(k_\rho\rts)
H_n^{(2)}(k_\rho\rb)
e^{jn(\phi_b-\phi_t)}
\right]dk_z,
s_i\in\{x,y\}.
\label{eq:dipole_position_derivative}
\end{multline}
The derivative inside the integral has the same range and angular structure as \eqref{eq:full_modal_kernel_derivative}, with $\ko$ replaced by the transverse wavenumber $k_\rho$. Substituting \eqref{eq:dipole_position_derivative} into \eqref{eq:full_modal_fim} gives the dipole-excitation FIM and the corresponding CRLB matrix and PEB in \eqref{eq:full_modal_crlb_peb}.}

\section{Numerical Results}
\label{Sec:NumericalResults}
\begin{figure}[t!]
    \centering
    \begin{subfigure}[b]{0.48\textwidth}
        \centering
        \pgfplotsset{every axis/.append style={
		font=\footnotesize,
		line width=1pt,
		legend style={font=\footnotesize, at={(0.99,0.99)}},legend cell align=left},
} \pgfplotsset{compat=1.13}
	\begin{tikzpicture}\begin{axis}[
xlabel near ticks,
grid=major,
xlabel={$\ar$ (m)},
ylabel={$|E_{z}|$ (dBV/m)},
width=0.99\linewidth,yticklabel style={/pgf/number format},
width=0.99\linewidth,
legend entries={Analytical\text{,} $\bt=-3~\text{cm} = \lambda/2$, 
	COMSOL Sim.\text{,} $\bt=-3~\text{cm} =\lambda/2$, Analytical\text{,} $\bt=-30$~cm,  
    COMSOL Sim.\text{,} $\bt=-30$~cm,
	},
 	log ticks with fixed point, 
ymin=29, ymax=36.5,
ylabel style={font=\footnotesize},
xlabel style={font=\footnotesize},
]
\addplot[red,solid,mark=diamond,mark repeat=20,mark size=3] table {Figures/EMReLine2Absd3axisX.dat};
\addplot[gray!50!black,mark=x,mark repeat=20,mark size=3,mark options=solid, only marks] table {Figures/EMReLine2AbsSimd3axisX.dat};
\addplot[blue,solid,mark=diamond,mark repeat=20,mark size=3] table {Figures/EMReLine2Absd30axisX.dat};
\addplot[orange!20!brown,mark=x,mark repeat=20,mark size=3,mark options=solid, only marks] table {Figures/EMReLine2AbsSimd30axisX.dat};

\end{axis}
\end{tikzpicture}
        \vspace{-0.2cm}
        \caption{Absolute value of $E_\text{tot}$}
        \vspace{0.2cm}
        \label{fig:AbsELine2}
    \end{subfigure}
    \hfill
    \begin{subfigure}[b]{0.48\textwidth}
        \centering
        \pgfplotsset{every axis/.append style={
		font=\footnotesize,
		line width=1pt,
		legend style={font=\footnotesize, at={(0.99,0.99)}},legend cell align=left},
} \pgfplotsset{compat=1.13}
	\begin{tikzpicture}\begin{axis}[
xlabel near ticks,
ylabel near ticks,
grid=major,
xlabel={$\ar$ (m)},
ylabel={$\operatorname{Re}\{E_z\}$ (V/m)},
width=0.99\linewidth,yticklabel style={/pgf/number format},
width=0.99\linewidth,
legend entries={Analytical\text{,} $\bt=-3~\text{cm} = \lambda/2$, COMSOL Sim.\text{,} $\bt=-3~\text{cm} =\lambda/2$, Analytical\text{,} $\bt=-30$~cm,  
    COMSOL Sim.\text{,} $\bt=-30$~cm,
	},
	ymin=-4000, ymax=5000,
ylabel style={font=\footnotesize},
xlabel style={font=\footnotesize},
]
\addplot[red,solid,mark=diamond,mark repeat=15,mark size=3] table {Figures/EMReLine2d3axisX.dat};
\addplot[gray!50!black,mark=x,mark repeat=15,mark size=3,mark options=solid, only marks] table {Figures/EMReLine2Simd3axisX.dat};
\addplot[blue,solid,mark=diamond,mark repeat=15,mark size=3] table {Figures/EMReLine2d30axisX.dat};
\addplot[orange!20!brown,mark=x,mark repeat=15,mark size=3,mark options=solid, only marks] table {Figures/EMReLine2Simd30axisX.dat};

\end{axis}
\end{tikzpicture}
        \vspace{-0.3cm}
        \caption{Real part of $E_\text{tot}$}
        \label{fig:RealELine2}
    \end{subfigure}
    \caption{Total received electric field along the x-axis (Line 2) with the target at the origin and $\br=0$, for $\bt \in \{3, 30\}$~cm: (a) absolute value and (b) real part.}
    \label{fig:CombinedELine2}
\end{figure}

\begin{figure}[t!]
    \centering
    \begin{subfigure}[b]{0.4\textwidth}
        \centering
        \includegraphics[width=0.99\linewidth]{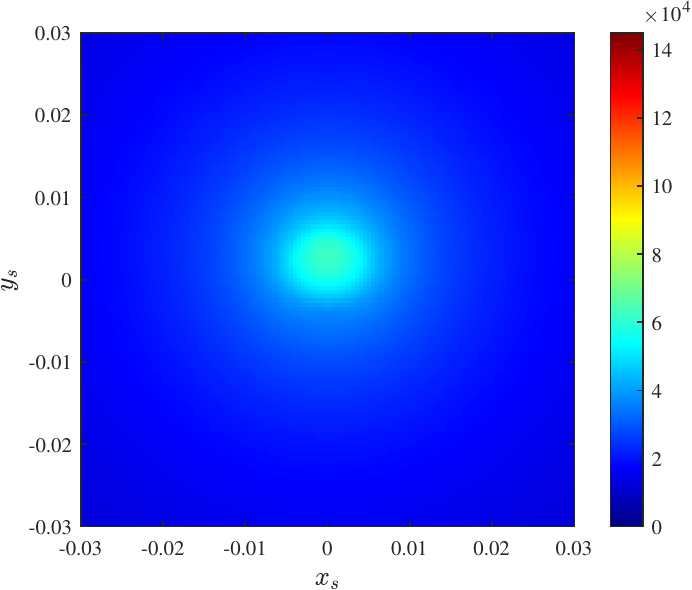}
        \caption{$\epsd=3.38$}
        \label{fig:Heatmaper3}
        \vspace{0.1cm}
    \end{subfigure}
    \hfill
    \begin{subfigure}[b]{0.4\textwidth}
        \centering
        \includegraphics[width=0.99\linewidth]{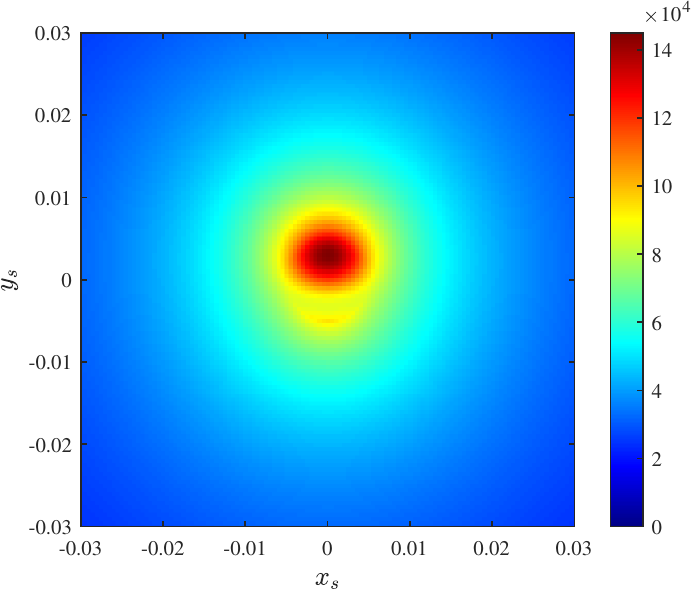}
        \caption{$\epsd=10.2$}
        \label{fig:Heatmaper10}
    \end{subfigure}
    \caption{{Absolute value of the scattered field for various receiver locations, with the target at the origin and a transmitting dipole at $(0, -0.03)$. (a) $\epsd=3.38$ and (b) $\epsd=10.2$.}}
    \label{fig:HeatmapField}
\end{figure}

Numerical results are presented and discussed for a single-cylinder scenario depicted in Fig.~\ref{fig:Config}, where the excitation is applied either through a line source or a dipole. In the following subsections, we present the numerical results for the analytically derived electric field and the sensing performance in terms of the \ac{CRLB} on target location estimation. { In the numerical results, we retain the modal orders $n=-3,\ldots,3$, which provide sufficient accuracy since the reconstruction error becomes negligibly small when more modes are included.}

\subsection{Accuracy of the Analytical Framework for Field Computation}

In Fig.~\ref{fig:ISAC_er_COMSOL}, we present the simulation setup used to validate our analytical results. A dielectric cylinder with a radius of ${a=0.5}$~cm (${d=1}$~cm) is placed at the center of the computational domain. The dielectric material of the target is characterized by a relative permittivity of $\epsd=3.38$. It is excited by infinitely long current filaments aligned along the $\hat{\mathbf{z}}$-axis at a frequency of $5$~GHz. The current filament is located at $(\at=0, \bt \in \{-3, -30\}$~cm).

{
The COMSOL validation uses the two-dimensional \emph{Electromagnetic Waves,
Frequency Domain} interface. The cylinder is embedded in a circular air domain
of radius $0.50$~m, surrounded by a concentric annular perfectly matched layer
(PML) extending from $0.50$ to $0.60$~m, and hence having a thickness of
$0.10$~m. The outer boundary behind the PML is terminated by the default
perfect-electric-conductor condition, while the standard electromagnetic
field-continuity conditions are enforced at the air--dielectric interface. A
free triangular mesh with a custom global size specification is used. The
maximum and minimum element sizes are $\lambda_0/10=0.006$~m and
$2.4\times10^{-5}$~m, respectively; the maximum element-growth rate is $1.1$,
the curvature factor is $0.2$, and the narrow-region resolution is $1$. The
resulting mesh contains $94{,}588$ domain elements and $1{,}184$ boundary
elements, and the frequency-domain system contains $66{,}373$ solved degrees
of freedom.}

The electric field (both its absolute value and real part) along Line 1 and Line 2, as depicted in Fig.~\ref{fig:ISAC_er_COMSOL}, is shown in Fig.~\ref{fig:CombinedELine1} and Fig.~\ref{fig:CombinedELine2}, respectively. It can be observed that the analytical results closely match the simulation outcomes obtained from COMSOL. In Fig.~\ref{fig:AbsELine1}, the absolute value of the electric field decreases with distance for both excitation positions, due to limited interference between the incident field (behind the cylinder) and the scattered field along Line 1. However, this behavior changes in Fig.~\ref{fig:AbsELine2} for $\bt=30$~cm, where the scattered field interferes with the incident field, leading to a fading effect. Additionally, our model accurately captures the phase of the electric field, as seen in Fig.~\ref{fig:RealELine1} and Fig.~\ref{fig:RealELine2}.

Unlike the previous figures that consider line excitation, Fig.~\ref{fig:HeatmapField} illustrates the case of dipole excitation, with the transmitting dipole positioned at $(\ar=0, \br=-0.03)$. The absolute value of the scattered field is shown for various receiver locations, considering two different values of the target's relative dielectric constant: $\epsd=10.2$ and $\epsd=3.38$, shown in Figs.~\ref{fig:Heatmaper10} and \ref{fig:Heatmaper3}, respectively.  As expected, it can be observed that the scattered field is stronger around the target when the relative dielectric constant is higher. This behavior is attributed to the increased contrast between the dielectric properties of the target and the surrounding medium (typically free space). A higher permittivity induces a greater discontinuity in the electric field at the material boundary, resulting in stronger scattering. Consequently, objects with higher permittivity values exhibit greater electromagnetic visibility, thereby facilitating more reliable detection.
\subsection{Sensing Performance Based on the Derived Field Model}

In this subsection, we present the numerical results for the ultimate sensing accuracy in estimating the target position, considering our model with two potential excitations (i.e., line current filament and dipole).

\begin{figure}
    \centering
    \includegraphics[width=\linewidth]{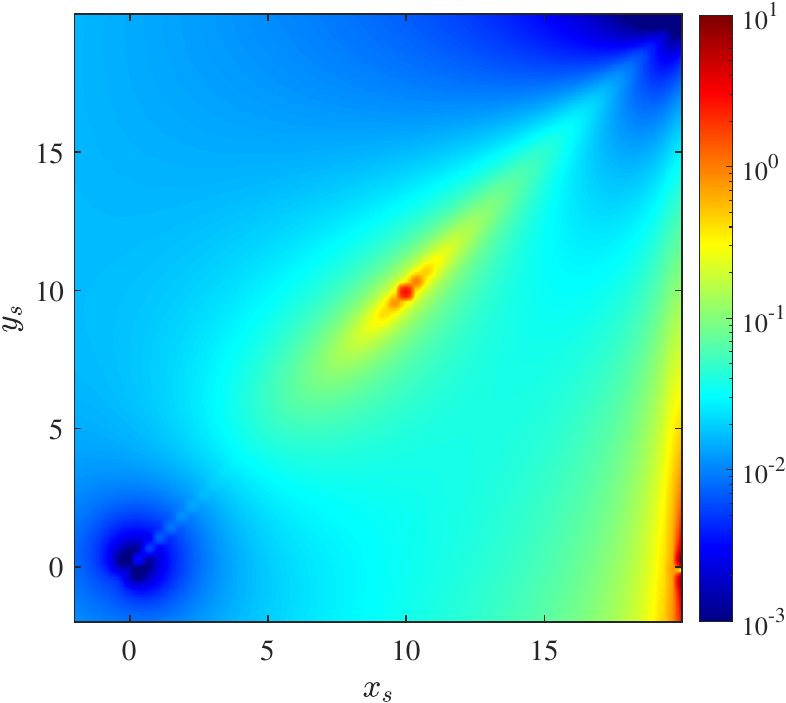}
    \caption{PEB in metres across various target locations for line excitation. The transmitter is at the origin, and the multi-antenna receiver is centred at $(20,20)$.}
    \label{fig:CRLBHMLine}
\end{figure}

Figure~\ref{fig:CRLBHMLine} presents the PEB computed from \eqref{eq:full_modal_crlb_peb} for a target with $\epsd=3.38$ at various spatial locations under line excitation. The transmit power is selected such that the scattered-field \ac{SNR} at the receiver-array centre is $40$~dB when the target is at the centre of the considered area. The transmitter is placed at the origin, and the receiver is an eight-element vertical ULA with half-wavelength inter-element spacing, centred at $(20,20)$. The sensing error is minimal in the vicinity
of the transmitter and/or receiver and becomes more significant
along the line connecting the two. The PEB becomes larger
along the Tx–Rx line because small lateral movements of the
target do not significantly affect the total travel distance or
the phase of the scattered field, resulting in measurements
that carry limited information about lateral displacement. Off
that line, even slight shifts alter the total path length (the
bistatic range), increasing the information content and thereby
lowering the PEB.

{
To complement the spatial PEB map, we evaluate the full-modal line-source model in a bistatic distance experiment with a fixed target at $\mathbf{s}_0=(0,0)$~m and a fixed transmitter at $(0,-0.03)$~m. The centre of an eight-element vertical ULA is moved along $[\ar,\br]=(R,0)$, while the inter-element spacing is fixed to $\lambda/2$. At $5$~GHz, the resulting array aperture is $D=(\Nb-1)\lambda/2=0.210$~m and the Fraunhofer distance is $d_{\rm F}=2D^2/\lambda=1.469$~m. The distance is swept over $R/d_{\rm F}\in[0.05,20]$, retaining the exact target-to-element distances $\rho_b$ throughout.

The noise variance is selected such that the average scattered-field SNR across the eight receiving antennas is $0$, $10$, or $20$~dB when the array centre is located at $R=d_{\rm F}$. For each reference SNR, the same noise variance is retained over the complete distance sweep, so that the effective received SNR varies naturally with the receiver-array distance. Each RMSE point is obtained from $1000$ independent noise realizations.

\begin{figure}[t!]
    \centering
    \includegraphics[width=\linewidth]{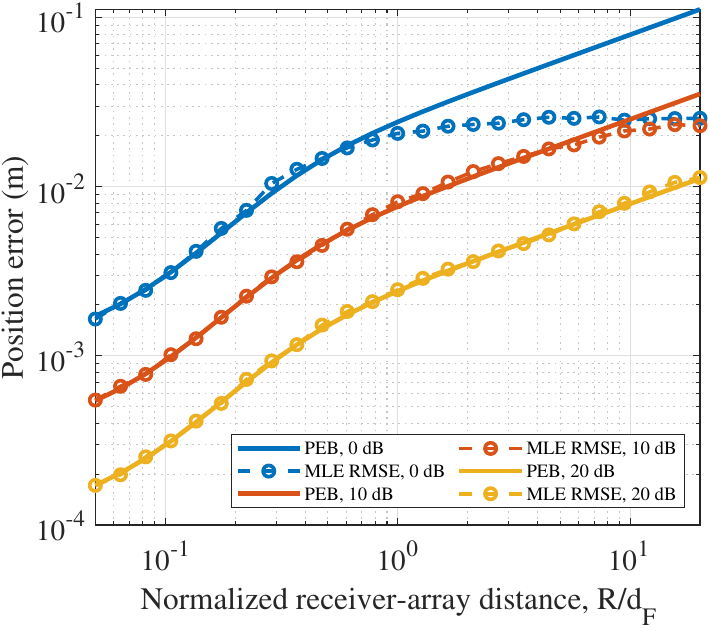}
    \caption{{PEB and MLE RMSE for the line-source model versus normalized receiver-array distance.}}
    \label{fig:line_mle_crlb_distance}
\end{figure}

Figure~\ref{fig:line_mle_crlb_distance} compares the PEB with the RMSE of a maximum-likelihood estimator. At $20$~dB, the simulated RMSE closely follows the PEB over the complete near-to-far distance range, while the $10$~dB result follows the bound over a broad range. For the $0$~dB reference setting, the departure becomes more visible at larger receiver-array distances because the scattered-field power decreases while the noise variance remains fixed, causing the effective received SNR to fall below its reference value. In this low-SNR threshold region, nonlinear estimators may exhibit outliers or boundary-dominated estimates and need not attain the local PEB~\cite[Ch.~2]{VanTrees:02}. The high-SNR agreement between the  estimator and the PEB verifies the internal consistency of the field derivatives and FIM.}

\begin{figure}[t!]
    \begin{subfigure}[b]{0.4\textwidth}
        \centering
\includegraphics[width=0.99\linewidth]{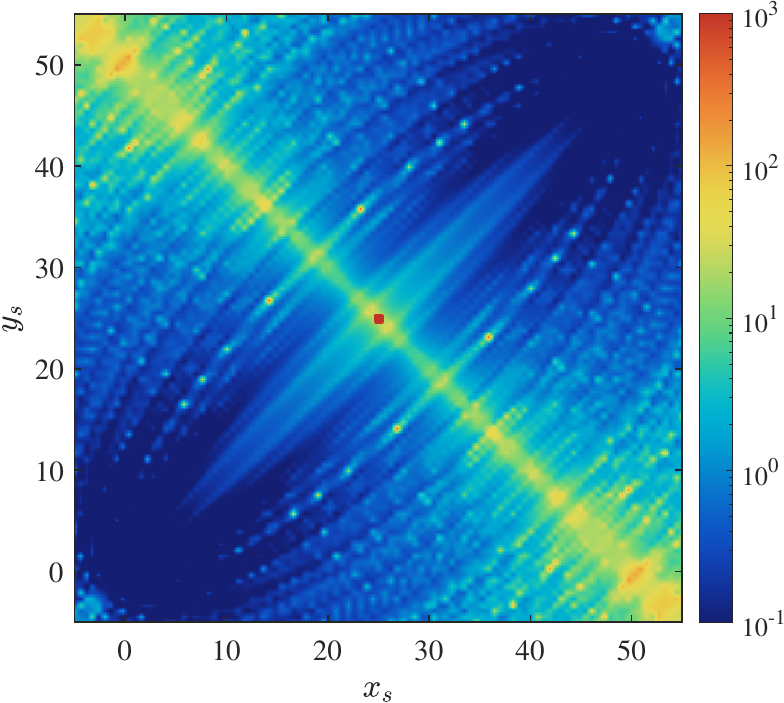}
        \caption{$\epsd=3.38$}
        \label{fig:CRLBDipoleeps3}
    \end{subfigure}
    \begin{subfigure}[b]{0.4\textwidth}
        \centering
        \includegraphics[width=0.99\linewidth]{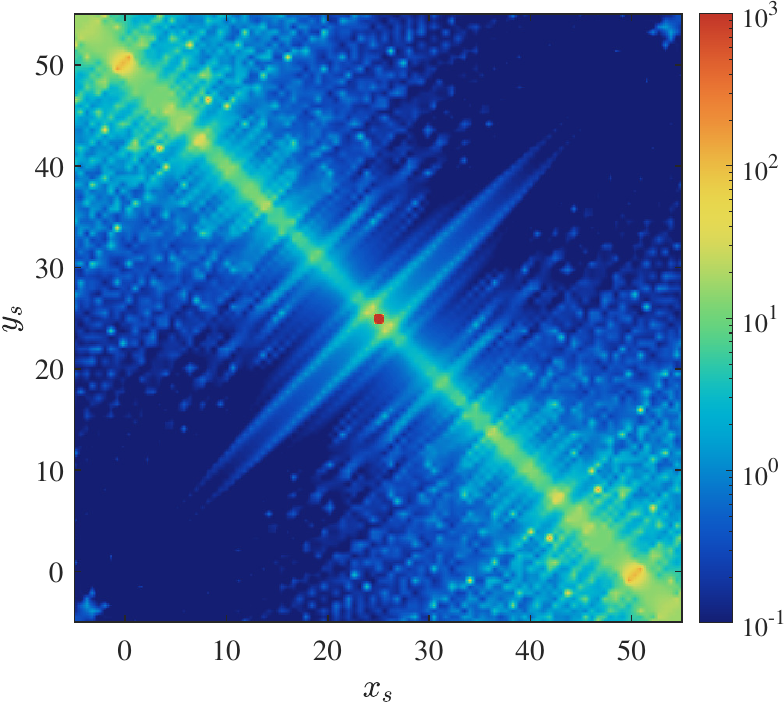}
        \caption{$\epsd=10.2$}
        \label{fig:CRLBDipoleeps10}
    \end{subfigure}
    \caption{PEB in metres across various target locations using the dipole EM model with the transmitter at the origin and a single receiving antenna at $(50,50)$, with $\epsd \in \{3.38,10.2\}$.}
    \label{fig:CRLBDipole}
\end{figure}
Figure~\ref{fig:CRLBDipole} shows the PEB for position estimation using the dipole-based EM model with a single receiving antenna. The electric-field noise variance is selected such that the scattered-field SNR at the receiving antenna is $40$~dB when the target with $\epsd=3.38$ is at the centre of the considered area. Comparing Figs.~\ref{fig:CRLBDipoleeps3} and \ref{fig:CRLBDipoleeps10}, we observe that the higher dielectric constant, i.e., $\epsd=10.2$, leads to a lower PEB, indicating improved detectability of a target that is more electrically visible compared with one with $\epsd=3.38$. Nevertheless, thanks to the electromagnetic model of the wave scattered by the target, it is still possible to detect and localize the target with reasonable accuracy. Also, we observe that the PEB is high on the line perpendicular to the Tx--Rx baseline, i.e., the locus of $\rho_{sr}=\rho_{ts}$. One reason is that, when $\rho_{sr}=\rho_{ts}$, some of the gradients with respect to the target position become small or vanish. This effect is more evident in the single-receiver case, whereas it is mitigated in Fig.~\ref{fig:CRLBHMLine} by the use of a multi-antenna receiver.

\begin{figure}[t!]
    \begin{subfigure}[b]{0.4\textwidth}
        \centering
\includegraphics[width=0.99\linewidth]{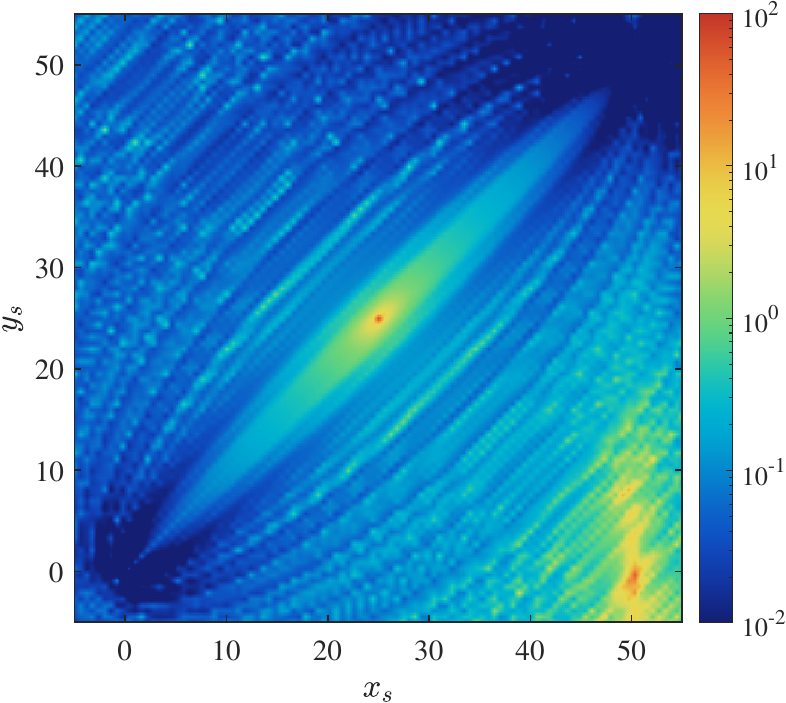}
        \caption{$\epsd=3.38$}
        \label{fig:CRLBDipoleeps3N4}
    \end{subfigure}
    \begin{subfigure}[b]{0.4\textwidth}
        \centering
        \includegraphics[width=0.99\linewidth]{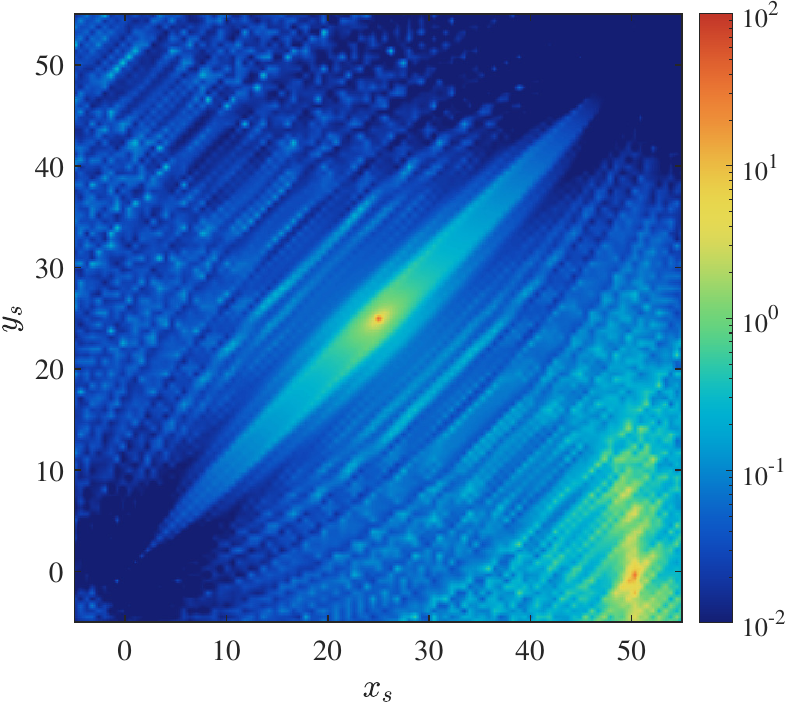}
        \caption{$\epsd=10.2$}
        \label{fig:CRLBDipoleeps10N4}
    \end{subfigure}
    \caption{PEB in metres across various target locations using the dipole EM model with the transmitter at the origin and a four-antenna receiver centred at $(50,50)$, with $\epsd \in \{3.38,10.2\}$.}
    \label{fig:CRLBDipoleN4}
\end{figure}

In this regard, Fig.~\ref{fig:CRLBDipoleN4} reports the PEB for the dipole case under the same setting as Fig.~\ref{fig:CRLBDipole}, but with a four-antenna vertical ULA with half-wavelength inter-element spacing. We observe similar behavior to the line-source case, even along the off-diagonal, i.e., the line perpendicular to the baseline. This can be attributed to the additional degrees of freedom obtained by observing the field at multiple antenna elements.

\subsection{Channel Capacity Evaluation}

\begin{figure*}[t!]
    \centering
    \includegraphics[width=\textwidth]{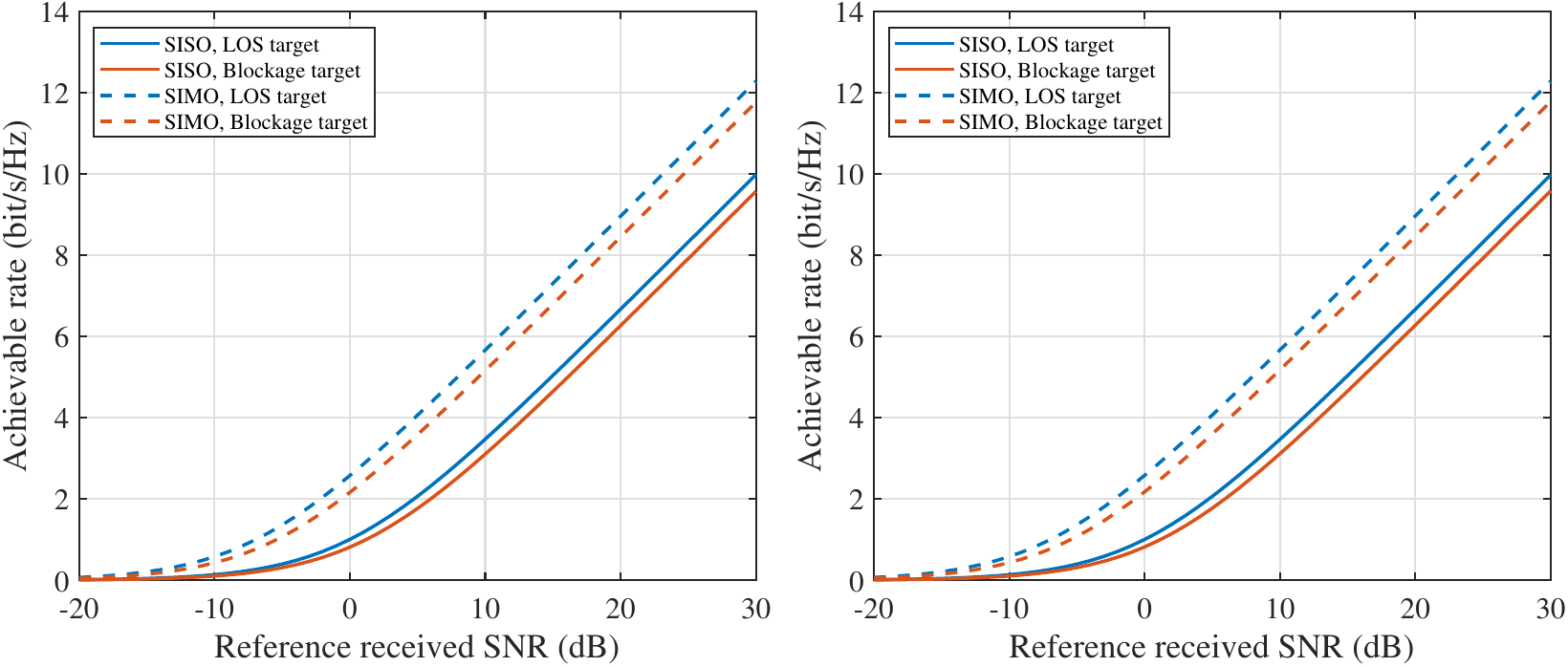}
    \caption{Achievable rate for SISO and five-element SIMO
    reception under LOS-target and blockage-target conditions. The
    horizontal axis is the no-target reference received SNR at the
    receiver-array centre. The left and right panels correspond to line-source
    and dipole excitation, respectively.}
    \label{fig:CapacityEvaluation}
\end{figure*}

{
To further illustrate the connection between the proposed EM channel model
and communication-side ISAC performance, we evaluate the achievable
capacity using the EM-derived receive-field response. For each excitation type
$q\in\{\mathrm{line},\mathrm{dip}\}$, we define the reference field as the
no-target field evaluated at the centre of the receive array,
\begin{equation}
    E_{\mathrm{ref}}^{(q)}
    \triangleq
    E_{\mathrm{no\mbox{-}target}}^{(q)}(\mathbf{r}_{\mathrm{c}}).
    \label{eq:capacity_reference_field}
\end{equation}
The resulting dimensionless channel coefficient at the $b$-th receive antenna
is
\begin{equation}
    h_b^{(q)}
    \triangleq
    \frac{E_{\mathrm{tot},b}^{(q)}}{E_{\mathrm{ref}}^{(q)}},
    \qquad b=1,\ldots,B,
\end{equation}
where $E_{\mathrm{tot},b}^{(q)}$ is the coherent direct-plus-scattered field
predicted by the corresponding full-modal EM model. The reference received
SNR is defined as
\begin{equation}
    \gamma_{\mathrm{ref}}^{(q)}
    \triangleq
    \frac{|E_{\mathrm{ref}}^{(q)}|^2}{\sigma_{E,q}^2},
\end{equation}
where $\sigma_{E,q}^2$ is the electric-field noise variance. With coherent
maximal-ratio combining, the output SNR and achievable rate are
\begin{equation}
    \gamma_{\mathrm{out}}^{(q)}
    =\gamma_{\mathrm{ref}}^{(q)}\|\mathbf{h}^{(q)}\|^2,
    \qquad
    C^{(q)}=\log_2\!\left(1+\gamma_{\mathrm{out}}^{(q)}\right).
    \label{eq:capacity_em_channel}
\end{equation}
Thus, the horizontal axis in Fig.~\ref{fig:CapacityEvaluation} is the
target-independent reference received SNR, while the target placement changes
$\|\mathbf{h}^{(q)}\|^2$ and hence the actual post-combining SNR and achievable
rate. In the SISO case, $\mathbf{h}^{(q)}$ reduces to a scalar.

The considered geometry is as follows. The transmitter is located at the
origin, while the SISO receiver is placed at $[5\lambda_0,5\lambda_0]$. For the
SIMO case, a five-element vertical receive array with half-wavelength spacing
is centred at the same location. Two target placements are considered: a LOS
configuration with the cylinder located at $[3\lambda_0,6\lambda_0]$, and an
NLOS/blockage configuration with the cylinder located at
$[3\lambda_0,3\lambda_0]$. The operating frequency is $5$ GHz, the cylinder
radius is $a=0.005$ m, and the relative permittivity is $\epsd=10.2$.
}

{
Figure~\ref{fig:CapacityEvaluation} shows the resulting achievable-rate curves.
The normalization gives unit no-target gain in the SISO case. For the
five-element SIMO receiver, the no-target gains are approximately $5.01$ and
$5.05$ for line-source and dipole excitation, respectively, corresponding to
the expected $7$~dB receive-array gain. The LOS target causes only a small
change for the selected geometry. By contrast, the blockage target reduces the
SISO normalized gains to $0.761$ and $0.767$ for line-source and dipole
excitation, respectively, corresponding to losses of approximately $1.19$ and
$1.15$~dB relative to the respective no-target reference fields. For SIMO, the
blockage-target gains are $3.48$ and $3.53$, compared with the corresponding
no-target gains of $5.01$ and $5.05$. The two excitation models therefore give
similar, but not identical, relative channel changes in this geometry. These
results connect the target-aware EM response to channel gain, post-combining
SNR, and achievable rate without treating the post-combining SNR itself as the
swept independent variable.
}

\section{Advantages and Open Challenges}
\label{sec:Challenges}
\subsection{Advantages}
The single target study presented here has highlighted the ability to localize small targets with weak dielectric contrast through an accurate channel model. 

It is thus evident that the proposed electromagnetic channel model offers several distinct advantages compared to: i) conventional full-wave numerical solvers typically used in propagation modeling and ii) simplified system-level models typically used in ISAC studies. 
On the one hand, standard EM solvers discretize the entire solution domain, resulting in a prohibitively large number of unknowns (exacerbated in multi-target scenarios). In contrast, the number of unknowns in our approach scales directly with the number of scatterers, such as dielectric cylinders, making it significantly more computationally efficient.
Moreover, while EM solvers require the careful implementation of radiation boundary conditions to ensure accuracy and stability in open-domain simulations, our formulation circumvents this need entirely by employing closed-form analytical expressions. This eliminates a key source of numerical complexity and reduces the dependency on meshing strategies or absorbing layers.
The availability of explicit closed-form solutions not only facilitates efficient computation but also renders the model inherently scalable and highly amenable to integration with stochastic geometry frameworks. As such, the proposed model serves as a foundational building block for future analytical studies of large-scale ISAC networks, enabling tractable performance evaluation and optimization across a wide range of deployment scenarios.
On the other hand, the electromagnetic channel model provides mathematically tractable formulas that can be harnessed to calculate closed-form expressions under different excitation conditions. This is particularly useful when the scattering from the target is weak due to low dielectric contrast (and sub-wavelength radius). Thanks to an accurate characterization of the scattered field, which is effectively as accurate as full-wave simulations, it is still possible to detect and localize a weakly scattering target. 
{
Compared with simplified sensing channel models commonly
used in ISAC studies, the proposed framework retains the
complex-valued electromagnetic response of the target,
including its amplitude, phase, and dependence on the
target parameters. Such physical consistency is important
for localization, since inaccuracies in the assumed channel
model may introduce model mismatch and affect estimation
performance \cite{AlexElzanaty:25,fortunati2017performance}.
This is particularly relevant to near-field ISAC, where
localization performance depends on the adopted propagation
model and estimation framework \cite{11150585}.
}

\subsection{Challenges}
The proposed electromagnetic channel model represents a key step forward in ISAC-oriented wireless propagation modeling, but several challenges must be addressed for broader application and real-world deployment. Modeling clusters of irregular, asymmetric scatterers requires handling complex near-field interactions efficiently, which is essential for realistic indoor and urban environments. Capturing wave behavior in more intricate settings—like coupled rooms or layered media—calls for incorporating coupling effects, edge diffraction, and hybrid modeling techniques.

Transitioning to full 3D vectorial models is also crucial to accurately represent polarization and field components in real-world geometries, though closed-form solutions remain difficult. Practical deployments demand accounting for variability in target position, orientation, and material properties, which suggests embedding stochastic elements into the model. Additionally, extending the near-field radar cross-section to multiple interacting targets introduces new challenges in interpreting scattered energy and defining suitable sensing metrics.

Scalability is another major concern, particularly for large or real-time systems. Reduced-order models or machine learning surrogates trained on electromagnetic data could bridge the gap between accuracy and feasibility. Solving these issues will enable more predictive, physics-based models that support both communication and precision sensing.

\section{Conclusion}
\label{sec:conclusion}
We have introduced a physics-based electromagnetic channel model for integrated sensing and communication (ISAC) in scattering environments. Based on wave theory, the model accounts for both line-source and dipole excitations and characterizes wave-object interactions across near- and far-field regimes. Unlike statistical or ray-based models, our approach captures full-wave effects, including polarization, material dispersion, and angular response, improving localization and sensing accuracy. We have derived the channel transfer function and quantified localization performance through closed-form expressions for the Cramér-Rao lower and position error bounds, validated against full-wave simulations. The model characterizes target-induced scattering and provides insight into field phase information often neglected in conventional RCS-based models. These results establish a foundation for electromagnetic information-theoretic analysis in ISAC scenarios and support more reliable channel modelling for 3GPP-like environments. Future work will extend this framework to multiple, potentially asymmetric scatterers and full 3D vectorial formulations, enabling richer propagation modeling for emerging wireless applications.

\appendices
\setlength{\abovedisplayskip}{4pt}
\setlength{\belowdisplayskip}{4pt}
\setlength{\abovedisplayshortskip}{3pt}
\setlength{\belowdisplayshortskip}{3pt}
\section{Boundary Conditions on Cylinders for Line-Source Excitation}
\label{apendix:boundrycondlinesource}

The scattering coefficients are determined by enforcing the
continuity of the tangential electric and magnetic fields
at the surface of the dielectric cylinder, $\rho=a$.
For line-source excitation, the electric field is related
to the auxiliary scalar potential by
$E_z=-j\omega\muo\psi$.
Since the cylinder is nonmagnetic, continuity of the
tangential electric field gives
\begin{equation}
    \psi_{\textrm{i}}(a^{+},\phi)
    +\psi_{\textrm{s}}(a^{+},\phi)
    =\psi_{\textrm{s}}(a^{-},\phi),
    \label{eq:econtfields}
\end{equation}
while continuity of the tangential magnetic field requires
\begin{equation}
    H_{\phi}^{\textrm{i}}(a^{+},\phi)
    +H_{\phi}^{\textrm{s}}(a^{+},\phi)
    =H_{\phi}^{\textrm{in}}(a^{-},\phi),
    \label{eq:mcontfields}
\end{equation}
where the superscripts $\textrm{i}$, $\textrm{s}$, and
$\textrm{in}$ denote the incident, scattered, and
interior fields, respectively. In accordance with
\eqref{eq.scatteredline}, $\psi_{\textrm{s}}(a^{-},\phi)$
represents the transmitted potential inside the cylinder.

Substituting \eqref{eq.incidentline} and
\eqref{eq.scatteredline} into \eqref{eq:econtfields},
and cancelling the common factor $1/(4j)$, yields
the first boundary condition.

The transverse distance between the transmitter
and a point on the cylinder surface is
$d(\phi)=\sqrt{\rts^{2}+a^{2}-2\rts a\cos(\phi-\phi')}$.
Therefore,
\begin{align}
 \sum_{n=-\infty}^{\infty}
 I_n J_n(\kd a)e^{jn(\phi-\phi')}
 =&\sum_{n=-\infty}^{\infty}
 I_n^{\textrm{s}}H_n^{(2)}(\ko a)
 e^{jn(\phi-\phi')}
 \nonumber\\
 &+\It H_0^{(2)}(\ko d(\phi)).
\end{align}
The first term on the right-hand side represents
the scattered contribution, whereas the second
term is the incident field generated by the line source.

Using the addition theorem for the Hankel function,
we have
\begin{align}
    \!\!H_0^{(2)}
    \left(\ko
    |\boldsymbol{\rho}_{\text{ts}}
    +\boldsymbol{\rho}|\right)
    &=
    \sum_{n=-\infty}^{\infty}
    H_n^{(2)}(\ko\rts)
    J_n(\ko\rho)
    e^{jn(\phi-\phi')},
    \label{eq:graf_line_addition_appendix}
\end{align}
which holds for
$|\boldsymbol{\rho}|<\rts$.
Since the line source is located outside the cylinder,
$a<\rts$, and the expansion is valid on its surface.
Substituting $\rho=a$ and equating the coefficients
of the independent angular harmonics gives
\begin{equation}
\label{eqn:bc1}
    I_n J_n(\kd a)
    =
    \It H_n^{(2)}(\ko\rts)J_n(\ko a)
    +I_n^{\textrm{s}}H_n^{(2)}(\ko a).
\end{equation}
Likewise, the magnetic field is related to the
auxiliary scalar potential through
$H_\phi=-\partial\psi/\partial\rho$.
Thus, the second boundary condition in
\eqref{eq:mcontfields} can be expressed as
\begin{equation}
\begin{aligned}
\left.H_\phi^{\textrm{in}}\right|_{\rho=a^-}
&=
-\left.
\frac{\partial\psi_{\textrm{s}}}{\partial\rho}
\right|_{\rho=a^-}
\\
&=
-\left.
\frac{\partial}{\partial\rho}
(\psi_{\textrm{i}}+\psi_{\textrm{s}})
\right|_{\rho=a^+}.
\end{aligned}
\end{equation}
Here, the incident auxiliary potential is
\begin{equation}
    \psi_{\textrm{i}}
    =
    \frac{\It}{4j}
    \sum_{n=-\infty}^{\infty}
    H_n^{(2)}(\ko\rts)
    J_n(\ko\rho)
    e^{jn(\phi-\phi')}.
\end{equation}
Substituting the incident, scattered, and interior
modal expansions into the magnetic-field boundary
condition, and again cancelling the common
normalization factor, yields
\begin{equation}
\label{eqn:bc2}
    \kd I_n J_n'(\kd a)
    =
    \ko\It H_n^{(2)}(\ko\rts)J_n'(\ko a)
    +
    \ko I_n^{\textrm{s}}H_n^{(2)\prime}(\ko a).
\end{equation}
Finally, arranging \eqref{eqn:bc1} and
\eqref{eqn:bc2} in matrix form gives
\eqref{eqn:FieldContFil}.

\section{Boundary conditions on cylinders for dipole excitation}
\label{appendix:bondatconddipole}
We exploit the continuity of the $z$ component of the tangential electric field across the cylinder surface as
\begin{equation}
    \left . E_z \right |_{\rho=a^{-}} = \left . E_z \right |_{\rho=a^{+}}, \,\,\, \textrm{for} \,\,\, 0 \leq \phi \leq 2 \pi, -\infty < z < \infty,
\end{equation}
which yields through (\ref{eqn:dEz}) \cite[Eq.~(9.2-2)]{balanis2012advanced}. 
\begin{align} \label{eqn:bcEs1}
& \left . \frac{1}{j \omega \epso \epsd } \left ( \frac{\partial^2}{\partial z^2} + \ko^2  \epsd \right ) \, d\psi_1^m \right |_{\rho=a} \nonumber \\ 
&= \left . \frac{1}{j \omega \epso } \left ( \frac{\partial^2}{\partial z^2} + \ko^2  \right ) \, \left ( d\psi^i + d\psi_s^m \right ) \right |_{\rho=a}.
\end{align}
By substituting \eqref{eq:fieldincidentdipole}, \eqref{eqn:dpsims}, and  \eqref{eqn:dpsimsi} into \eqref{eqn:bcEs1}, for the $n^{\text{th}}$ scattering mode we obtain
\begin{multline}\label{eqn:EzCont}
 -k^2_{\rho}\, H_{n}^{(2)} (k_{\rho} a ) \, I^{s}_n+\frac{k^2_{\rho d}}{\epsd}  J_n \left ( k_{\mathsf{\rho d}} a \right ) \, I_n \\ 
 = k^2_{\rho}\,  I^t H_{n}^{(2)} (k_{\rho} \, \rho_{ts} ) J_n \left ( k_{\rho} a \right ),
\end{multline}
which is the first equation in \eqref{eqn:FieldCont}. 

We now proceed to impose the continuity of the $\phi$ component of the tangential electric field across the cylinder surface: 
\begin{equation}
    \left . E_{\phi} \right |_{\rho=a^{-}} = \left . E_{\phi} \right |_{\rho=a^{+}}, \,\,\, \textrm{for} \,\,\, 0 \leq \phi \leq 2 \pi, -\infty < z < \infty,
\end{equation}
which yields \cite[Eq.~(9.2-2)]{balanis2012advanced} 
\begin{align} \label{eqn:bcEs2}
& \left . \frac{1}{j \omega \epso \epsd} \frac{\partial^2}{\rho \partial \phi \partial z} \, d\psi_1^m + \frac{\partial}{\partial \rho} d\psi_1^e \right |_{\rho=a} \nonumber \\ 
&= \left . \frac{1}{j \omega \epso} \frac{\partial^2}{\rho \partial \phi \partial z} \, \left ( d\psi_i +  d\psi_s^m \right ) + \frac{\partial}{\partial \rho} d\psi_s^e \right |_{\rho=a}.
\end{align}
By substituting \eqref{eq:fieldincidentdipole} -- \eqref{eqn:dpsimsi} into \eqref{eqn:bcEs2}, for the $n^{\text{th}}$ scattering mode we obtain
\begin{multline}\label{eqn:EphiCont}
 {\frac{j k_z}{\omega \epso} \, \frac{n}{a} \, H_{n}^{(2)} (k_{\rho} a )} \,
 I^{s}_n -
 k_{\rho} \, H_{n}^{(2)'} {\left(k_{\rho}a\right)} \, K^{s}_n - \\
 \frac{j k_z}{\omega \epso \epsd} \, \frac{n}{a} \, J_n \left ( k_{\rho d} a \right ) \, I_n + 
 k_{\rho d} \, J_n^{'} \left ( k_{\mathsf{\rho d}} a \right ) \, K_n \\
 = {- \frac{j k_z}{\omega \epso} \, \frac{n}{a} \, I^t H_{n}^{(2)} (k_{\rho} \, \rho_{ts} ) J_n \left ( k_{\rho} a \right )},
\end{multline}
which is the second equation in \eqref{eqn:FieldCont}. 

The continuity of magnetic fields is further imposed across the cylinder surface. 
The $z$ component of the tangential magnetic field yields 
\begin{equation}
    \left . H_z \right |_{\rho=a^{-}} = \left . H_z \right |_{\rho=a^{+}}, \,\,\, \textrm{for} \,\,\, 0 \leq \phi \leq 2 \pi, -\infty < z < \infty,
\end{equation}
which gives \cite[Eq.~(9.2-2)]{balanis2012advanced}
\begin{align} \label{eqn:bcHs1}
& \left . \frac{1}{j \omega \muo } \left ( \frac{\partial^2}{\partial z^2} + \ko^2  \epsd \right ) \, d\psi_1^e \right |_{\rho=a} \nonumber \\ 
&= \left . \frac{1}{j \omega \muo } \left ( \frac{\partial^2}{\partial z^2} + \ko^2  \right ) \, d\psi_s^e \right |_{\rho=a}.
\end{align}
By substituting \eqref{eq:fieldincidentdipole}, \eqref{eq:dfieldsoutside}, and \eqref{eq:dfieldinsidedipole} into \eqref{eqn:bcHs1}, for the $n^{\text{th}}$ scattering mode we obtain
\begin{equation}\label{eqn:HzCont}
 k^2_{\rho} \, H_{n}^{(2)} (k_{\rho} a ) \, K^{s}_n - k^2_{\rho d} \, J_n \left ( k_{\mathsf{\rho d}} a \right ) K_n = 0, 
\end{equation}
which is the third equation in \eqref{eqn:FieldCont}. 

Finally, the continuity of the $\phi$ component of the tangential magnetic field yields
\begin{equation}
    \left . H_{\phi} \right |_{\rho=a^{-}} = \left . H_{\phi} \right |_{\rho=a^{+}}, \,\,\, \textrm{for} \,\,\, 0 \leq \phi \leq 2 \pi, -\infty < z < \infty,
\end{equation}
which gives \cite[Eq.~(9.2-2)]{balanis2012advanced}
\begin{align} \label{eqn:bcHs2}
& \left . \frac{1}{j \omega \muo} \frac{\partial^2}{\rho \partial \phi \partial z} \, d\psi_1^e - \frac{\partial}{\partial \rho} d\psi_1^m \right |_{\rho=a} \nonumber \\ 
&= \left . \frac{1}{j \omega \muo} \frac{\partial^2}{\rho \partial \phi \partial z} \, d\psi_s^e - \frac{\partial}{\partial \rho} \left ( d\psi_i + {d\psi_s^m} \right ) \right |_{\rho=a}.
\end{align}
By substituting \eqref{eq:fieldincidentdipole} -- \eqref{eqn:dpsimsi} into \eqref{eqn:bcHs2}, for the $n^{\text{th}}$ scattering mode we obtain
\begin{equation}\label{eqn:HphiCont}
{
\begin{split}
&-k_{\rho} H_{n}^{(2)'}(k_{\rho}a) I_n^s
-\frac{j k_z}{\omega\muo}\frac{n}{a}H_n^{(2)}(k_{\rho}a)K_n^s \\
&\quad +k_{\rho d}J_n'(k_{\rho d}a)I_n
+\frac{j k_z}{\omega\muo}\frac{n}{a}J_n(k_{\rho d}a)K_n \\
&=k_{\rho}I^tH_n^{(2)}(k_{\rho}\rho_{ts})J_n'(k_{\rho}a),
\end{split}
}
\end{equation}
which is the fourth equation in \eqref{eqn:FieldCont}. 

\section{Symmetry of modal coefficients}
From (\ref{eqn:FieldCont})-(\ref{eqn:H2recurr}) it can be concluded that $K^{s}_n (k_z)$ and $K_n (k_z)$ are odd functions of $k_z$, or
\begin{eqnarray}
    K^{s}_n (-k_z) & = & -K^{s}_n (k_z),\\
    {K_n (-k_z)} & {=} & {-K_n (k_z)}.
\end{eqnarray}
Moreover, we have 
\begin{eqnarray}
    I^{s}_n (-k_z) & = & I^{s}_n (k_z),\\
    {I_n (-k_z)} & {=} & {I_n (k_z)}.
\end{eqnarray}
Finally, by using the integer-order identities $J_{-n}(\cdot)=(-1)^nJ_n(\cdot)$ and $H^{(2)}_{-n}(\cdot)=(-1)^nH^{(2)}_n(\cdot)$, together with the source-dependent factor in the right-hand side of (\ref{eqn:FieldCont}), the negative-order coefficients satisfy
\begin{eqnarray}
    {K^{s}_{-n} (k_z)} & {=} & {(-1)^{n+1}K^{s}_n (k_z)},\\
    {K_{-n} (k_z)} & {=} & {(-1)^{n+1}K_n (k_z)}.
\end{eqnarray}
and 
\begin{eqnarray}
    {I^{s}_{-n} (k_z)} & {=} & {(-1)^nI^{s}_n (k_z)},\\
    {I_{-n} (k_z)} & {=} & {(-1)^nI_n (k_z)}.
\end{eqnarray}
\section{Incident Electric Field}
The $z$ component of the incident electric field is found by Faraday's law
\begin{equation}\label{eqn:Ezi}
 E^i_z = \frac{1}{j \omega \epso} \, \nabla \times \textbf{H}^i \cdot \hat{\bf z},    
\end{equation}
where the vector incident magnetic field is obtained from the vector potential $\textbf{A}$
\begin{equation}\label{eqn:Hi}
\begin{split}
 \textbf{H}^i = \nabla \times \textbf{A} = \frac{I^t dl}{4 \pi} \, \frac{\partial}{\partial \rtr} \left ( \frac{e^{-j \ko r}}{\rtr} \right ) \, \hat{\rtr} \times \hat{\bf z} \\
 = I^t dl \, \frac{e^{-j \ko r}}{4 \pi \rtr} \left (  j \ko + \frac{1}{r} \right ) \, \sin \theta \, \hat{\phi}.
\end{split}
\end{equation}
Given that from (\ref{eqn:Ezi}) 
\begin{equation}
 \left . E^i_z \right |_{z=0} = \left . - E^i_{\theta} \right |_{\theta=\frac{\pi}{2}} = \left . \frac{1}{j \omega \epso} \, \frac{1}{r \sin \theta} \, \frac{\partial}{\partial r} \left ( r \sin \theta H_{\phi} \right ) \right |_{\theta = \frac{\pi}{2}}, 
\end{equation}
using (\ref{eqn:Hi}) yields 
\begin{equation}
    \left . E^i_z \right |_{z=0} = \frac{I^t dl}{j \omega \epso} \, \frac{e^{-j \ko r}}{4 \pi} \left (  \frac{\ko^2}{r} - j \frac{\ko}{r^2} - \frac{1}{r^3} \right ),
\end{equation}
whose rearrangement gives (\ref{eqn:Ezi_final}).

\ifCLASSOPTIONcaptionsoff
  \newpage
\fi

\begin{IEEEbiography}[{\includegraphics[width=1in,height=1.25in,clip,keepaspectratio]{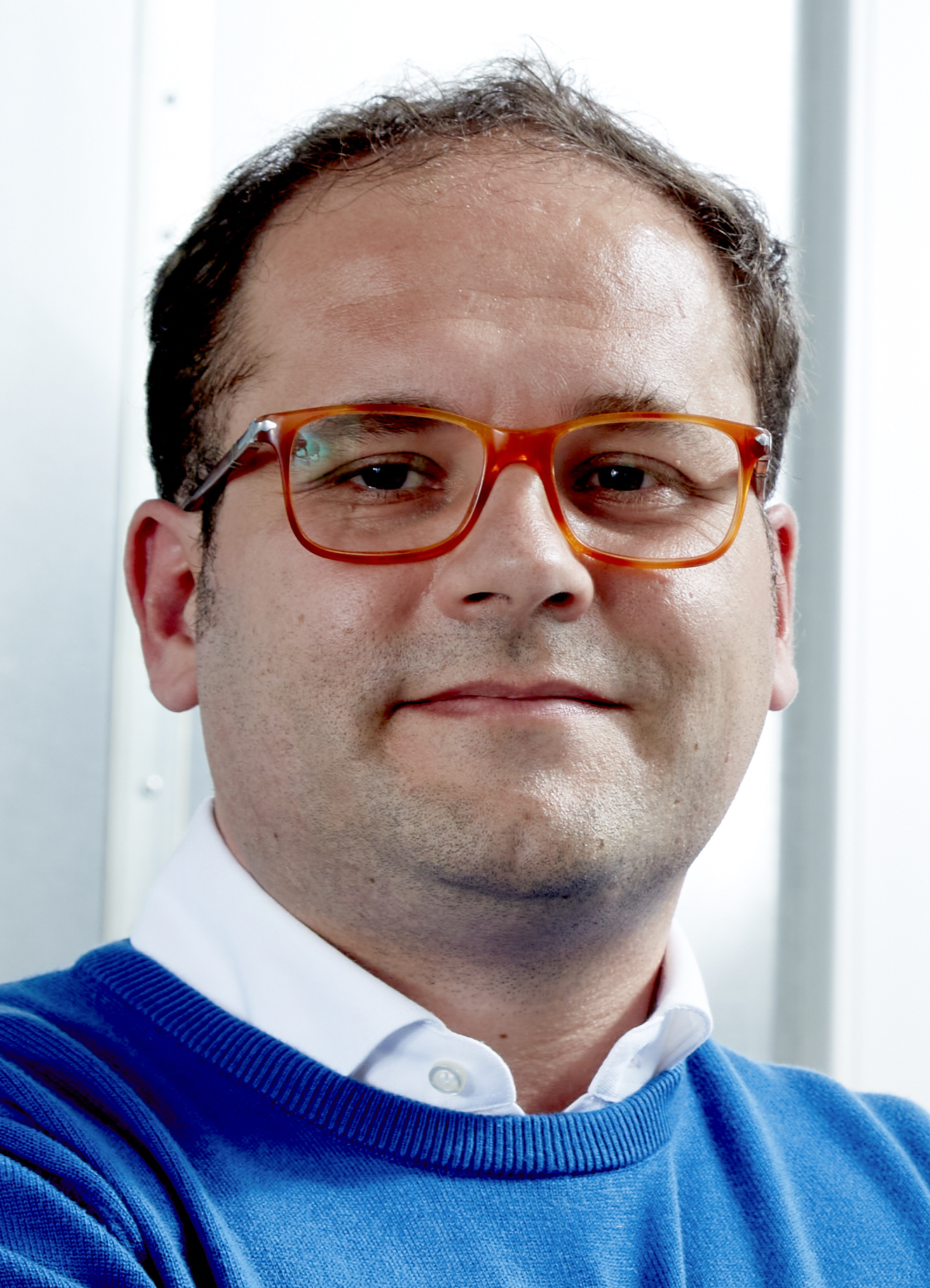}}]{Gabriele Gradoni}
(Member, IEEE) was a Professor of applied mathematics and electromagnetics engineering with the School of Mathematical Sciences and the George Green Institute for Electromagnetics Research, University of Nottingham, U.K. From 2020 to 2024, he was a Royal Society Industry Fellow with British Telecom, U.K. He is currently a Professor of wireless and satellite communications with the 6G Innovation Centre, Institute for Communication Systems, University of Surrey, Guildford, U.K. He is also a Visiting Fellow with the Computer Laboratory, University of Cambridge, U.K.; and an Adjunct Professor with the University of Illinois at Urbana--Champaign, USA. He has participated in several national and European projects, including HASC, NEMF21, and RISE-6G. His research interests include physics-based channel modeling for wireless communications, electromagnetic information theory, and quantum computational electromagnetics.
\end{IEEEbiography}
\vspace{-1.5cm}
\begin{IEEEbiography}[{\includegraphics[width=1in,height=1.25in,clip,keepaspectratio]{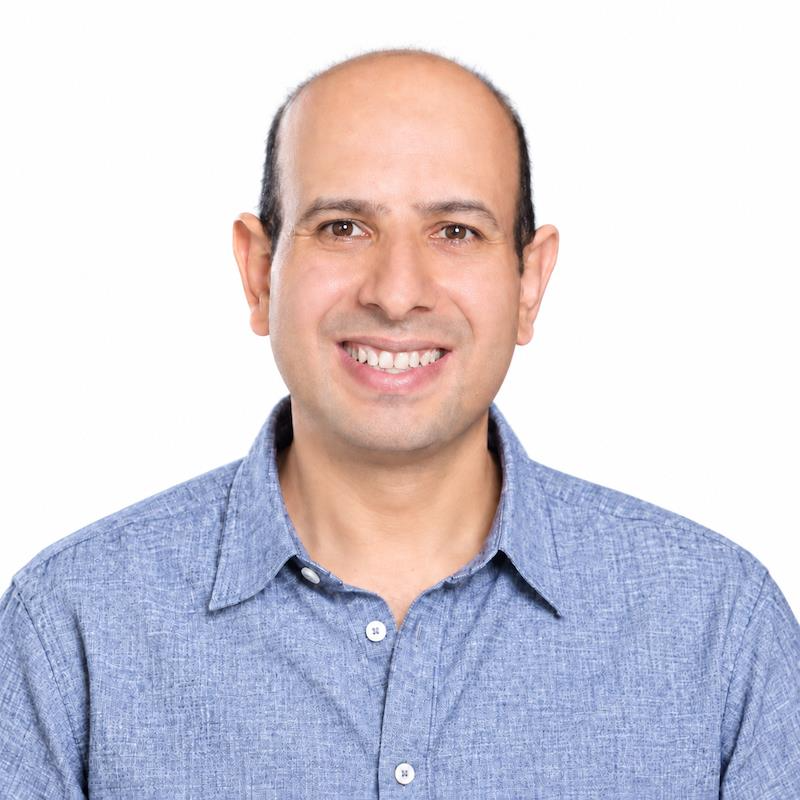}}]{Ahmed Elzanaty}
(Senior Member, IEEE) received the Ph.D. degree (Hons.) in electronics, telecommunications, and information technology from the University of Bologna, Italy, in 2018. He is currently a Lecturer (an Assistant Professor) with the Institute for Communication Systems, University of Surrey, U.K. He has participated in several national and European projects, including TUDOR, CHEDDAR, and EuroCPS. His research interests include the design and performance analysis of wireless communications and localization systems. He served as the Secretary for the Radio Communications Committee within the IEEE Communications Society.
\end{IEEEbiography}
\vspace{-1.5cm}
\begin{IEEEbiography}[{\includegraphics[width=1in,height=1.25in,clip,keepaspectratio]{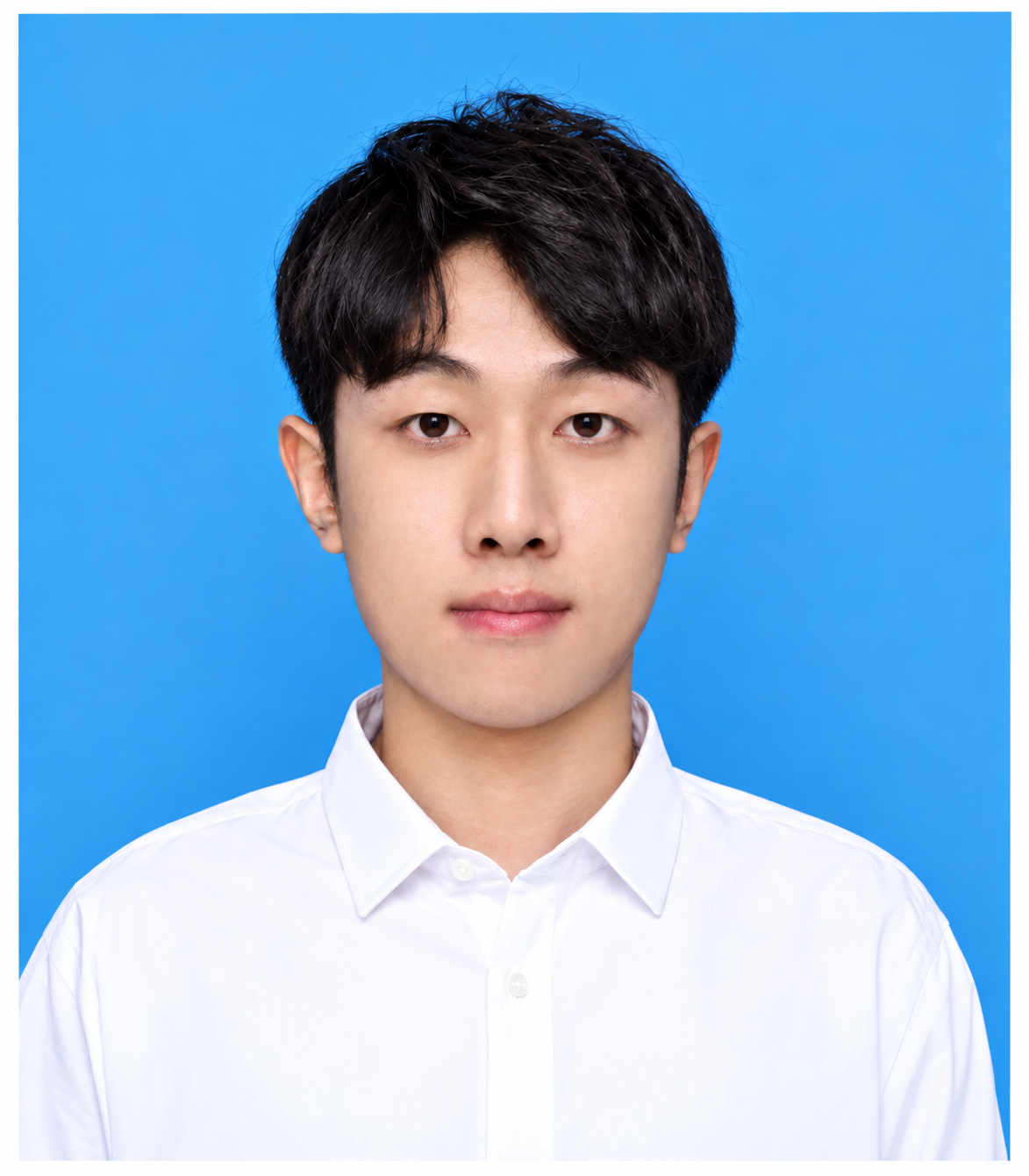}}]{Yadong Yang}
received the B.S. and M.S. degrees from Beijing University of Posts and Telecommunications, Beijing, China. He is currently pursuing the Ph.D. degree with the 6G Innovation Centre, Institute for Communication Systems, University of Surrey, U.K. His research interests include channel modeling, signal processing, and integrated sensing and communications.
\end{IEEEbiography}
\vspace{-1.5cm}
\begin{IEEEbiography}[{\includegraphics[width=1in,height=1.25in,clip,keepaspectratio]{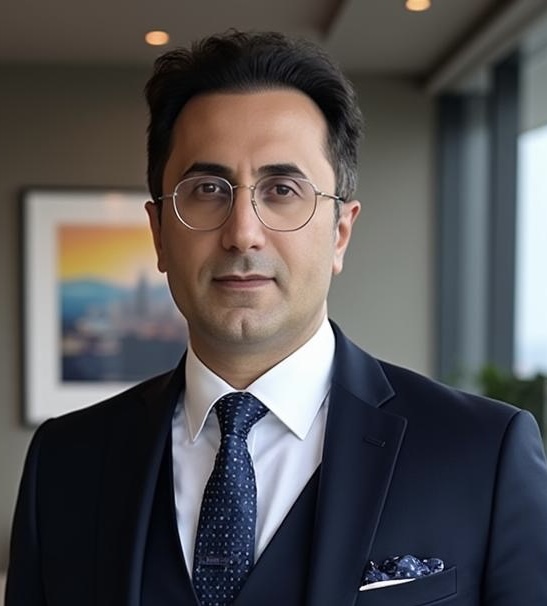}}]{Mohsen Khalily}
(Senior Member, IEEE) is currently an Associate Professor in antennas and propagation and the Head of the Surface Electromagnetics Laboratory (SEMLAB), Institute for Communication Systems, University of Surrey, U.K. He has published four book chapters and almost 200 academic articles in international peer-reviewed journals and conference proceedings. He has been the principal investigator on research grants totaling more than \$4.5 million in the field of surface electromagnetics. He is the leading rapporteur for the work item on implementation and practical considerations of the RIS in the dedicated Industry Specification Group within the ETSI. His research interests include surface electromagnetics, metamaterials, mmWave and THz antennas, and propagation modeling. He serves as an Associate Editor for IEEE TRANSACTIONS ON ANTENNAS AND PROPAGATION, IEEE ANTENNAS AND WIRELESS PROPAGATION LETTERS, and Scientific Reports (Nature Index).
\end{IEEEbiography}
\vspace{-1.5cm}
\begin{IEEEbiography}[{\includegraphics[width=1in,height=1.25in,clip,keepaspectratio]{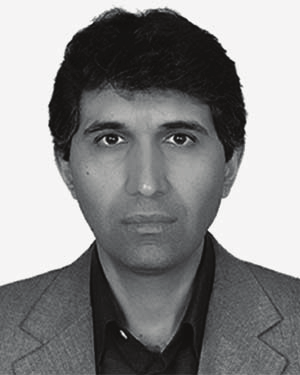}}]{Mahmoud Shahabadi}
(Senior Member, IEEE) received the B.Sc. and M.Sc. degrees in electrical engineering from the University of Tehran, Tehran, Iran, in 1988 and 1991, respectively, and the Ph.D. degree in electrical engineering from the Technische Universität Hamburg, Hamburg, Germany, in 1998. Since 1998, he has been with the School of Electrical and Computer Engineering, University of Tehran, where he is currently a Full Professor and the Founder and the Director of the Photonics Research Laboratory. He is currently a Visiting Professor with the University of Surrey. His current research interests include antenna engineering, terahertz engineering, and photonics. He served as the Vice-Chair and the Chair for the IEEE Iran Section from 2013 to 2014 and from 2015 to 2016, respectively.
\end{IEEEbiography}

\begin{IEEEbiography}[{\includegraphics[width=1in,height=1.25in,clip,keepaspectratio]{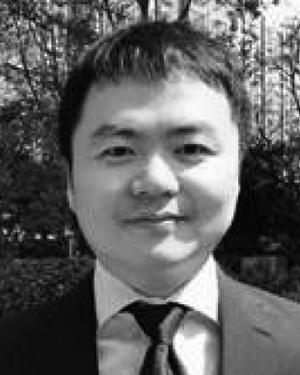}}]{Fan Wang}
received the B.S. and Ph.D. degrees in electronic engineering from Zhejiang University, Hangzhou, China. He joined Huawei Technologies Company Ltd., Shanghai, China, in 2008. From 2015 to 2016, he was with Huawei Technologies Sweden AB, Kista, Sweden. From 2019 to 2021, he was with Huawei Technologies Company Ltd., Reading, U.K. He has contributed to 3G/4G/5G physical-layer standardization work and holds numerous patents. His current research interests include B5G/6G wireless communications and massive machine-type communications.
\end{IEEEbiography}

\begin{IEEEbiography}[{\includegraphics[width=1in,height=1.25in,clip,keepaspectratio]{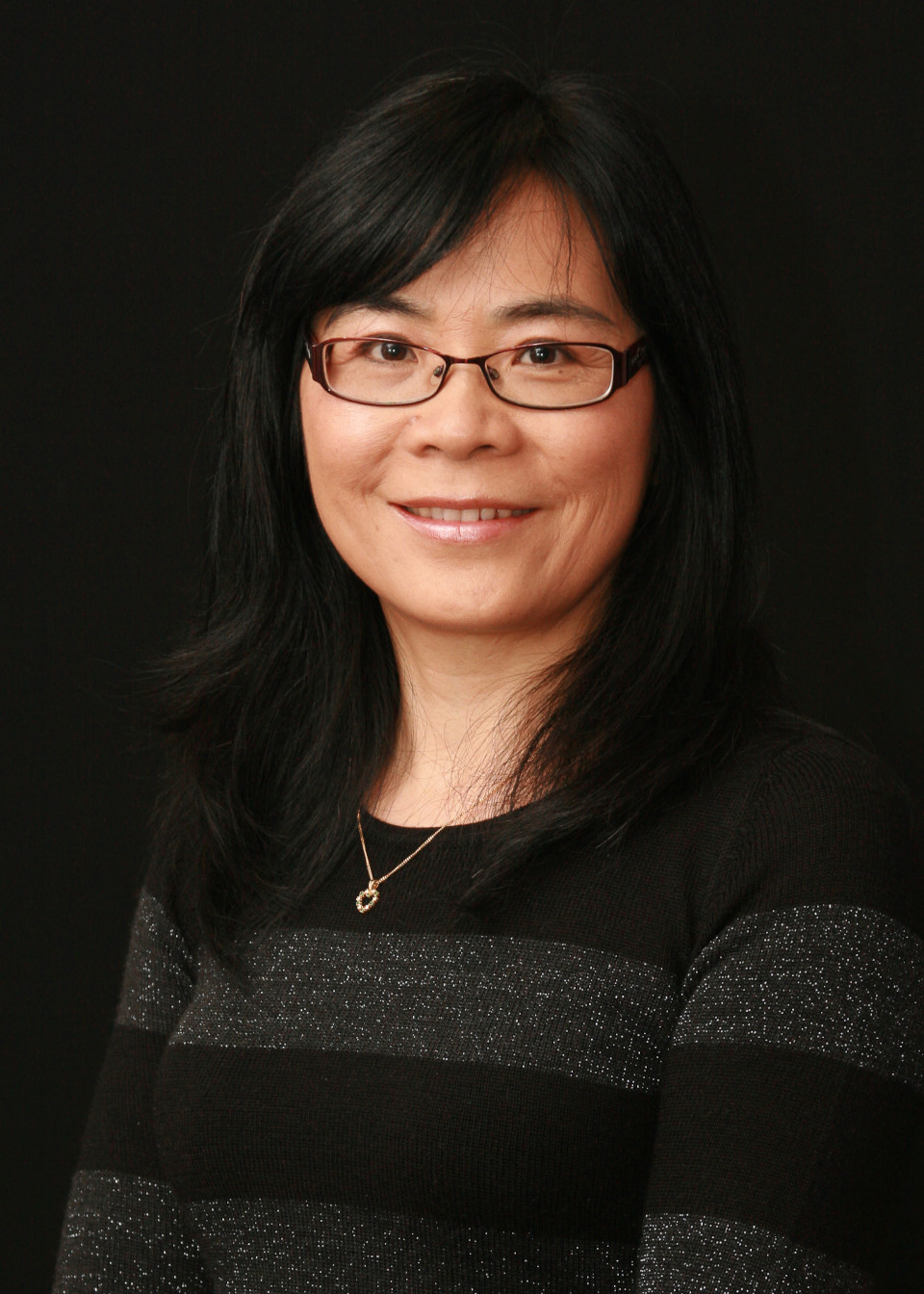}}]{Peiying Zhu}
(Fellow, IEEE) is currently leading 6G wireless research and standardization with Huawei Technologies Company Ltd. The focus of her research is advanced radio access technologies. She is actively involved in 3GPP and IEEE 802 standards development. She regularly gives talks and participates in panel discussions on 5G/6G vision and enabling technologies. She led the team that contributed significantly to 5G technologies and standardization. Many technologies developed by the team have been adopted into 5G standards and implemented in 5G products. Prior to joining Huawei in 2009, she was a Nortel Fellow and the Director of advanced wireless access technology with the Nortel Wireless Technology Laboratory. She led the team and pioneered research and prototyping on MIMO-OFDM and multi-hop relay, which were adopted into LTE standards and 4G products. She is the Senior Vice President of wireless research, a Huawei Fellow, and a fellow of Canadian Academy of Engineering.
\end{IEEEbiography}

\begin{IEEEbiography}[{\includegraphics[width=1in,height=1.25in,clip,keepaspectratio]{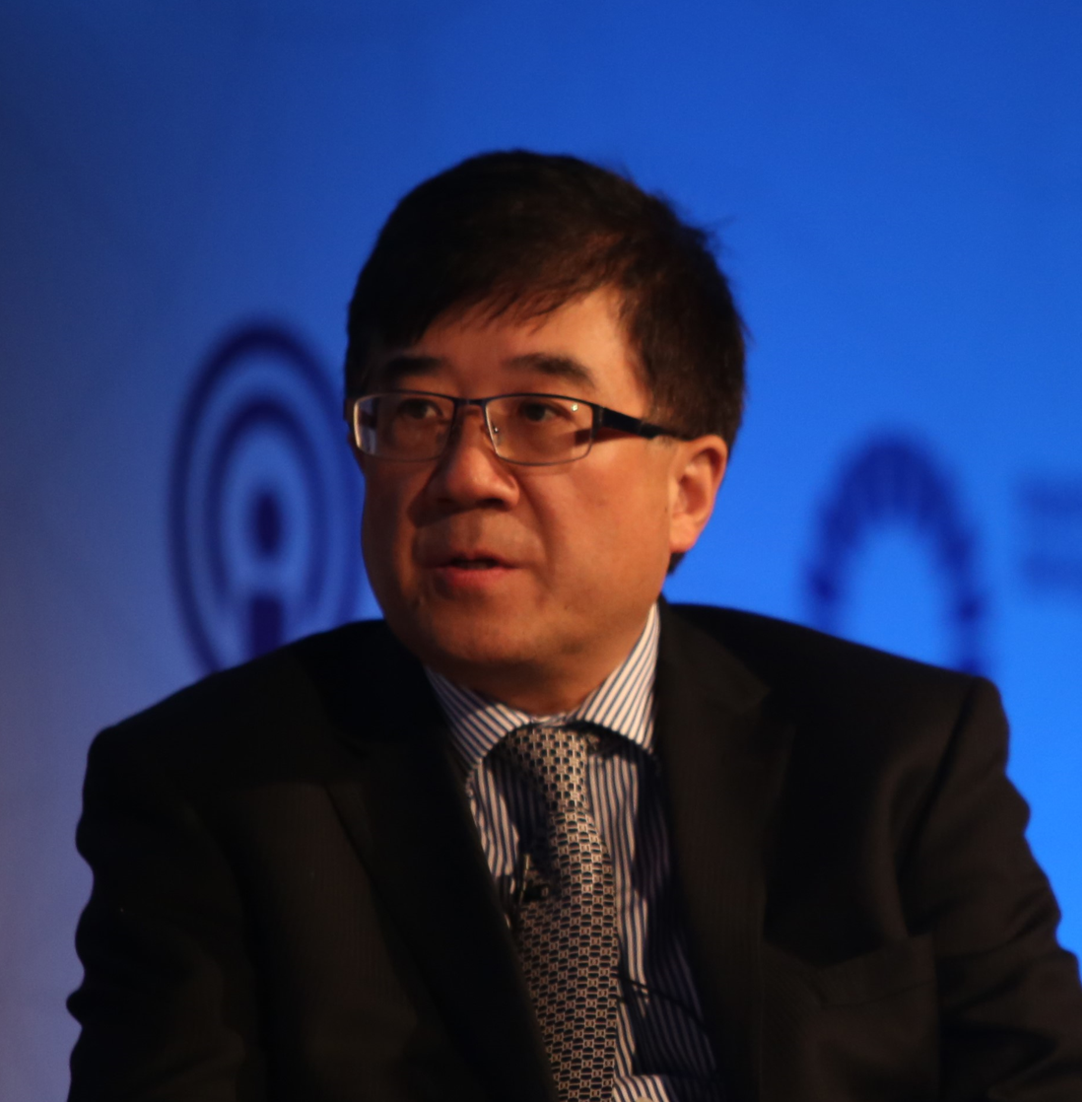}}]{Wen Tong}
(Fellow, IEEE) received the B.S. degree from the Department of Radio Engineering, Nanjing Institute of Technology, Nanjing, China, in 1984, and the M.Sc. and Ph.D. degrees in electrical engineering from Concordia University, Montreal, QC, Canada, in 1986 and 1993, respectively. In 2011, he became the Head of Communications Technologies Labs, Huawei Technologies Company Ltd., where he is currently the Huawei 5G Chief Scientist and leads Huawei's ten-year-long 5G wireless technologies research and development. For the past three decades, he has pioneered fundamental technologies from 1G to 5G wireless with more than 450 granted U.S. patents. He received the IEEE Communications Society Distinguished Industry Leader Award for ``Pioneering Technical Contributions and Leadership in the Mobile Communications Industry and Innovation in 5G Mobile Communications Technology'' in 2018. He also received the R. A. Fessenden Medal.
\end{IEEEbiography}

\begin{IEEEbiography}[{\includegraphics[width=1in,height=1.25in,clip,keepaspectratio]{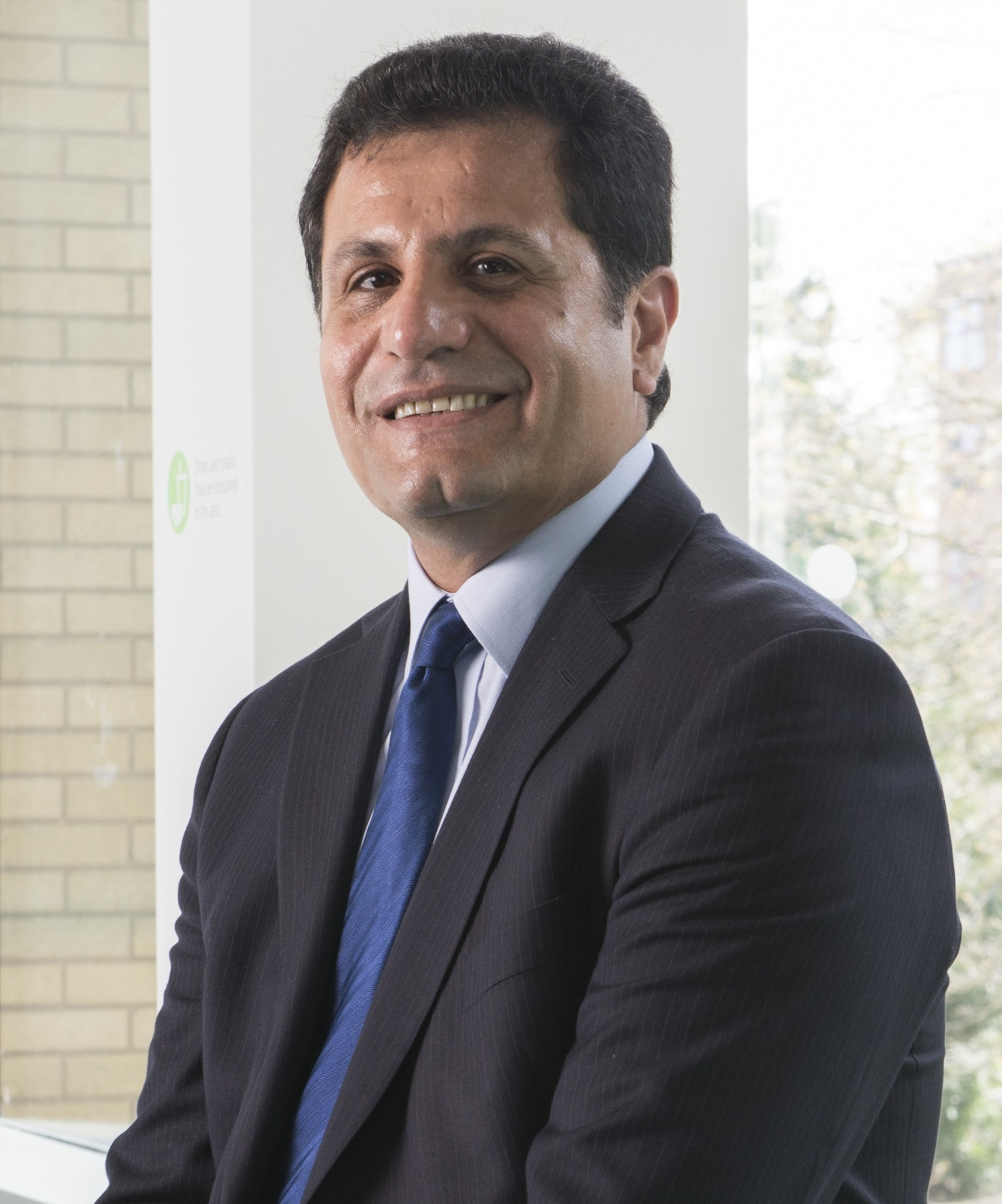}}]{Rahim Tafazolli}
(Fellow, IEEE) is currently a Professor of mobile and satellite communications, a Regius Professor of electronic engineering, and the Founder and the Director of the 5G and 6G Innovation Centre, Institute for Communication Systems, University of Surrey, U.K. He has been active in research for more than 30 years and has authored or co-authored more than 1200 research articles and books. He was elected a fellow of the Royal Academy of Engineering in 2020. He is also a fellow of IET and the Wireless World Research Forum. He is a Technical Advisor to many mobile companies and has lectured, chaired, and delivered keynote speeches at numerous IEEE workshops and conferences. He is nationally and internationally known in the field of mobile communications. He was honored as a Commander of the Order of the British Empire in recognition of his international contributions to mobile research more than the past 30 years.
\end{IEEEbiography}

\vfill

\vspace{-0.3cm}
\begin{thebibliography}{26}
\bibitem{liu2020joint}
F. Liu, C. Masouros, A. P. Petropulu, H. Griffiths, and L. Hanzo, ``Joint radar and communication design: Applications, state-of-the-art, and the road ahead,'' IEEE Trans. Commun., vol. 68, no. 6, pp. 3834--3862, Jun. 2020.

\bibitem{MasourosTWC:18}
F. Liu, C. Masouros, A. Li, H. Sun, and L. Hanzo, ``MU-MIMO communications with MIMO radar: From co-existence to joint transmission,'' IEEE Trans. Wireless Commun., vol. 17, no. 4, pp. 2755--2770, Apr. 2018.

\bibitem{ElzanatyAnnaFrans:21}
A. Elzanaty, A. Guerra, F. Guidi, and M.-S. Alouini, ``Reconfigurable intelligent surfaces for localization: Position and orientation error bounds,'' IEEE Trans. Signal Process., vol. 69, pp. 5386--5402, 2021.

\bibitem{ElzanatyMismatch:24}
A. Elzanaty, J. Liu, A. Guerra, F. Guidi, Y. Ma, and R. Tafazolli, ``Near and far field model mismatch: Implications on 6G communications, localization, and sensing,'' IEEE Internet Things Mag., vol. 7, no. 5, pp. 120--126, Sep. 2024.

\bibitem{AlexElzanaty:25}
A. M. Kuzminskiy, A. Elzanaty, G. Gradoni, F. Wang, and R. Tafazolli, ``Near-field localization with physics-compliant electromagnetic model: Algorithms and model mismatch analysis,'' IEEE Internet Things, vol. 12, no. 15, pp. 30533--30547, Aug. 2025.

\bibitem{luo2024channel}
C. Luo, A. Tang, F. Gao, J. Liu, and X. Wang, ``Channel modeling framework for both communications and bistatic sensing under 3GPP standard,'' IEEE J. Sel. Areas Sensors, vol. 1, pp. 166--176, 2024.

\bibitem{zhang2024cluster}
Z. Zhang et al., ``A cluster-based statistical channel model for integrated sensing and communication channels,'' IEEE Trans. Wireless Commun., vol. 23, no. 9, pp. 11597--11611, Sep. 2024.

\bibitem{zhang2026channel}
Z. Zhang et al., ``Channel semantic characterization for integrated sensing and communication scenarios: From measurements to modeling,'' IEEE Trans. Wireless Commun., vol. 25, pp. 13073--13089, 2026.

\bibitem{fugen2006capability}
T. Fügen, J. Maurer, T. Kayser, and W. Wiesbeck, ``Capability of 3-D ray tracing for defining parameter sets for the specification of future mobile communication systems,'' IEEE Trans. Antennas Propag., vol. 54, no. 11, pp. 3129--3139, Nov. 2006.

\bibitem{Reichardt2011}
L. Reichardt, J. Maurer, T. Fugen, and T. Zwick, ``Virtual drive: A complete V2X communication and radar system simulator for optimization of multiple antenna systems,'' Proc. IEEE, vol. 99, no. 7, pp. 1295--1310, Jul. 2011.

\bibitem{Zhang2021}
F. Zhang, C. Zhou, Y. Li, G. Xia, J. Zhu, and Z. Zhao, ``A GPU acceleration algorithm for urban electromagnetic environments,'' in Proc. 13th Int. Symp. Antennas, Propag. EM Theory (ISAPE), vol. 1, Dec. 2021, pp. 1--3.

\bibitem{miller2000communicating}
D. A. B. Miller, ``Communicating with waves between volumes: Evaluating orthogonal spatial channels and limits on coupling strengths,'' Appl. Opt., vol. 39, no. 11, pp. 1681--1699, 2000.

\bibitem{dardari2020communicating}
D. Dardari, ``Communicating with large intelligent surfaces: Fundamental limits and models,'' IEEE J. Sel. Areas Commun., vol. 38, no. 11, pp. 2526--2537, Nov. 2020.

\bibitem{pizzo2023wide}
A. Pizzo, A. Lozano, S. Rangan, and T. L. Marzetta, ``Wide-aperture MIMO via reflection off a smooth surface,'' IEEE Trans. Wireless Commun., vol. 22, no. 8, pp. 5229--5239, Aug. 2023.

\bibitem{pizzo2022spatial}
A. Pizzo, L. Sanguinetti, and T. L. Marzetta, ``Spatial characterization of electromagnetic random channels,'' IEEE Open J. Commun. Soc., vol. 3, pp. 847--866, 2022.

\bibitem{pizzo2022fourier}
A. Pizzo, L. Sanguinetti, and T. L. Marzetta, ``Fourier plane-wave series expansion for holographic MIMO communications,'' IEEE Trans. Wireless Commun., vol. 21, no. 9, pp. 6890--6905, Sep. 2022.

\bibitem{chew2001fast}
W. C. Chew, J.-M. Jin, E. Michielssen, and J. Song, Fast and Efficient Algorithms in Computational Electromagnetics Norwood, MA, USA: Artech House, 2001.

\bibitem{chew1993huygens}
W. C. Chew and C.-C. Lu, ``The use of Huygens' equivalence principle for solving the volume integral equation of scattering,'' IEEE Trans. Antennas Propag., vol. 41, no. 7, pp. 897--904, Jul. 1993.

\bibitem{he2020cascaded}
Z.-Q. He and X. Yuan, ``Cascaded channel estimation for large intelligent metasurface assisted massive MIMO,'' IEEE Wireless Commun. Lett., vol. 9, no. 2, pp. 210--214, Feb. 2020.

\bibitem{yee14numerical}
K. S. Yee, ``Numerical solution of initial boundary value problems involving Maxwell's equation,'' IEEE Trans. Antennas Propag., vol. AP-14, no. 3, pp. 302--307, May 1966.

\bibitem{jin2015finite}
J.-M. Jin, The Finite Element Method. Hoboken, NJ, USA: Wiley, 2015.

\bibitem{Harrington2001}
R. F. Harrington, ``Cylindrical wave functions,'' in Time-Harmonic Electromagnetic Fields. Hoboken, NJ, USA: Wiley, 2001, pp. 198--263.

\bibitem{balanis2012advanced}
C. A. Balanis, Advanced Engineering Electromagnetics. Hoboken, NJ, USA: Wiley, 2012.

\bibitem{VanTrees:02}
H. L. Van Trees, Classical Detection and Estimation Theory. Hoboken, NJ, USA: Wiley, 2002, ch. 2, pp. 19--165.

\bibitem{fortunati2017performance}
S. Fortunati, F. Gini, M. S. Greco, and C. D. Richmond, ``Performance bounds for parameter estimation under misspecified models: Fundamental findings and applications,'' IEEE Signal Process. Mag., vol. 34, no. 6, pp. 142--157, Nov. 2017.

\bibitem{11150585}
M. Rahal, A. Elzanaty, M. Mirmohseni, Y. Ma, and R. Tafazolli, ``Near-field wideband ISAC: Estimator design and performance analysis,'' IEEE Trans. Aerosp. Electron. Syst., vol. 62, pp. 1--14, 2026.

\end{thebibliography}
\end{document}